\documentclass[12pt]{iopart}

\RequirePackage{fix-cm}
\RequirePackage{xspace}
\usepackage{booktabs}
\usepackage{graphicx}
\usepackage{tabularx}
\usepackage{subfig}
\newsavebox{\measurebox}
\usepackage{multirow}
\usepackage{array}
\usepackage{epstopdf}
\usepackage{lineno}
\usepackage{rotating}
\usepackage{hyperref}

\expandafter\let\csname equation*\endcsname\relax
\expandafter\let\csname endequation*\endcsname\relax
\usepackage{amsmath}
\DeclareMathOperator\erf{erfc}
\usepackage{color}

\usepackage{preprintcover}  
\PreprintCoverPaperTitle{Measurement of the cross-section of high transverse momentum 
vector bosons reconstructed as single jets and studies of jet substructure 
in $pp$ collisions at  ${\sqrt{s}}$ = 7\,TeV with the ATLAS detector}  
\PreprintIdNumber{CERN-PH-EP-2014-123}  
\PreprintCoverAbstract{This paper presents a measurement of the cross-section for high transverse momentum $W$ and $Z$ bosons 
produced in $pp$ collisions and decaying to all-hadronic final states. The data used in the analysis were 
recorded by the ATLAS detector at the CERN Large Hadron Collider at a centre-of-mass energy of 
$\sqrt{s}=7~\rm TeV$ and correspond to an integrated luminosity of $4.6~\rm fb^{-1}$. The measurement 
is performed by reconstructing the boosted $W$ or $Z$ bosons in single jets. The reconstructed jet mass 
is used to identify the $W$ and $Z$ bosons, and a jet substructure method based on energy cluster 
information in the jet centre-of-mass frame is used to suppress the large multi-jet background. The 
cross-section for events with a hadronically decaying $W$ or $Z$ boson, with transverse momentum 
$p_{\rm T}>320\,{\rm GeV}$ and 
pseudorapidity $|\eta|<1.9$, is measured to be $\sigma_{W+Z}= 8.5 \pm 1.7$~pb and is compared to 
the next-to-leading-order calculations. 
The selected events are further used to study jet grooming techniques.}  
\PreprintJournalName{New J. Phys.}  

\begin{document}

\title[]{Measurement of the cross-section of high transverse momentum vector bosons 
reconstructed as single jets and studies of jet substructure  in $pp$ 
collisions at  ${\sqrt{s}}$ = 7\,TeV with the ATLAS detector}

\author{The ATLAS collaboration}
\date{Received: date / Accepted: date}

\def\WZ {\ensuremath{W/Z}\xspace}
\def\WplusZ {\ensuremath{W+Z}\xspace}
\def\pp {\ensuremath{pp}\xspace}
\def\mjet {\ensuremath{m_{\rm jet}}\xspace}
\def\PYTHIAEIGHT {{PYTHIA 8.153}\xspace}
\def\PYTHIASIX {{PYTHIA 6.426}\xspace}
\def\HERWIG {{HERWIG 6.520}\xspace}
\def\HERWIGPP {{HERWIG++ 2.6.3}\xspace}
\def\POWHEG {{POWHEG 1.0 (patch 4)}\xspace}
\def\JIMMY {{JIMMY 4.31}\xspace}

\def\pt{\ensuremath{p_{\mathrm{T}}}} 
\def\pT{\ensuremath{p_{\mathrm{T}}}} 
\def\alphas{\ensuremath{\alpha_{\mathrm{S}}}} 

\def\stat{\mbox{$\;$(stat.)}}
\def\syst{\mbox{$\;$(syst.)}}
\def\TeV{\ifmmode {\mathrm{\ Te\kern -0.1em V}}\else
                   \textrm{Te\kern -0.1em V}\fi}%
\def\GeV{\ifmmode {\mathrm{\ Ge\kern -0.1em V}}\else
                   \textrm{Ge\kern -0.1em V}\fi}%
\def\MeV{\ifmmode {\mathrm{\ Me\kern -0.1em V}}\else
                   \textrm{Me\kern -0.1em V}\fi}%
\def\keV{\ifmmode {\mathrm{\ ke\kern -0.1em V}}\else
                   \textrm{ke\kern -0.1em V}\fi}%
\def\eV{\ifmmode  {\mathrm{\ e\kern -0.1em V}}\else
                   \textrm{e\kern -0.1em V}\fi}%
\let\tev=\TeV
\let\gev=\GeV
\let\mev=\MeV
\let\kev=\keV
\let\ev=\eV

\begin{abstract} 
This paper presents a measurement of the cross-section for high transverse momentum $W$ and $Z$ bosons 
produced in \pp collisions and decaying to all-hadronic final states. The data used in the analysis were 
recorded by the ATLAS detector at the CERN Large Hadron Collider at a centre-of-mass energy of 
$\sqrt{s}=7~\rm TeV$ and correspond to an integrated luminosity of $4.6~\rm fb^{-1}$. The measurement 
is performed by reconstructing the boosted $W$ or $Z$ bosons in single jets. The reconstructed jet mass 
is used to identify the $W$ and $Z$ bosons, and a jet substructure method based on energy cluster 
information in the jet centre-of-mass frame is used to suppress the large multi-jet background. The 
cross-section for events with a hadronically decaying $W$ or $Z$ boson, with transverse momentum $\pt>320\,\gev$ and 
pseudorapidity $|\eta|<1.9$, is measured to be $\sigma_{W+Z}= 8.5 \pm 1.7$~pb and is compared to 
next-to-leading-order calculations. 
The selected events are further used to study jet grooming techniques.
\end{abstract}


\section{Introduction}\label{intro}

Many theories beyond the Standard Model (SM) predict new particles with masses at the~\tev\
scale. Some of these heavy resonances
can decay to final states with $W$ or $Z$ bosons. Because the 
masses of the $W$ and $Z$ bosons are an order of magnitude below that of their hypothetical
parent states, such decays  appear highly boosted in the laboratory
frame. The hadronic decay products  may be
so collimated that they appear as single jets in the detector (hereafter referred to as \WZ \emph{jets}). The ability to
recognize and reconstruct the $W$ and $Z$ bosons from such jets is important in extending
the search sensitivity for new phenomena with the ATLAS detector at the Large Hadron Collider (LHC).

Because the $W$ and $Z$ masses are small compared to the centre-of-mass energy of 7~\tev\ 
at the LHC, the $W$ and $Z$ bosons produced in SM processes can also be 
highly boosted in the detector.  An important first step in the study of boosted \WZ jets 
is to demonstrate that they can be measured reliably using the ATLAS detector. 
Jets arising from the strong interactions of quarks and gluons
(hereafter referred to as \emph{QCD jets}) have production cross-sections many orders of magnitude 
greater than those of $W$ and $Z$ bosons; they are the dominant background to such a measurement 
and constitute the main difficulty.    

In this paper, a measurement of the cross-section of hadronically decaying $W$ or $Z$ bosons 
with transverse momentum $\pT>320\,\gev$  and pseudorapidity $|\eta| < 1.9$ produced in \pp collisions at a 
centre-of-mass energy of 7~\tev\ is presented.\footnote{ATLAS uses a right-handed coordinate 
system with its origin at the nominal interaction point (IP) in the centre of the detector and 
the $z$-axis along the beam pipe. The $x$-axis points from the IP to the centre of the LHC ring, 
and the $y$-axis points upward. Cylindrical coordinates $(r,\phi)$ are used in the transverse 
plane, $\phi$ being the azimuthal angle around the beam pipe. The pseudorapidity is defined in 
terms of the polar angle $\theta$ as $\eta=-\ln\tan(\theta/2)$.} The measurement is based on 
the invariant mass distribution of the reconstructed boosted \WZ jet candidates. The decay modes 
considered are $W\to q\bar{q}^\prime$ and $Z\to q\bar{q}$, where $q, q^\prime = u, c, d, s$ 
or $b$. In order to suppress the copious QCD jet background, a novel selection 
method~\cite{Chen:2011ah} based on jet substructure is implemented.  Because of the limited 
resolution for the jet mass, the measurement reported is for the sum of $W$ and $Z$ cross-sections, 
denoted by the \WplusZ cross-section
\begin{eqnarray*}
 \sigma_{W+Z}&=&\sigma_W(\pt>320 \gev,|\eta|<1.9)\times \mathcal{B}(W\to q\bar{q}^\prime)\\
 &+&\sigma_Z(\pt>320 \gev,|\eta|<1.9)\times \mathcal{B}(Z\to q\bar{q}),
\end{eqnarray*}
where $\sigma$ is the production cross section and $\mathcal{B}$ is the decay branching fraction.
Previously, $W$ and $Z$ cross-sections have 
been measured up to $\pT=300\gev$ using the leptonic decay 
modes~\cite{Aad:2012en, Aad:2013ysa, Chatrchyan:2011wt}. The $Z$ cross-section for $\pT > 200 \gev$
has also been measured in the hadronic decay mode $Z\to b\bar{b}$~\cite{Aad:2014bla}. 

The jet sample enriched in $W$ and $Z$ bosons that decay hadronically, obtained in this analysis, 
is used to study the performance of several jet grooming 
techniques~\cite{ATLAS:2012am,Aad:2013gja,Chatrchyan:2012ku} designed to reduce the effects of 
soft QCD radiation and multiple \pp interactions per bunch crossing (pileup) on jet mass measurements.

\section{The ATLAS detector}
The ATLAS detector \cite{Aad:2008zzm}  at the LHC nearly covers the entire solid angle around the 
interaction region. It consists of an inner tracking detector comprising a silicon pixel detector, a 
silicon microstrip detector, and a transition radiation tracker, providing tracking capability 
within the pseudorapidity range $ |\eta| < 2.5$. The inner tracking detector  is surrounded by a thin 
superconducting solenoid providing a $2\,\rm T$ axial magnetic field and by a calorimeter system placed 
immediately outside the solenoid. 
The electromagnetic calorimeters use liquid argon as the active detector medium with lead absorbers, 
and are divided into one barrel
($|\eta|<1.475$) and two end-cap components ($1.375<|\eta|<3.2$). The technology used for
the hadronic calorimeters varies with $\eta$. In the barrel region ($|\eta|<1.7$), the detector is
made of scintillator tiles with steel absorbers. In the end-cap region ($1.5<|\eta|<3.2$)), the
detector uses liquid argon and copper. A forward calorimeter consisting of liquid argon and
tungsten/copper absorbers has both electromagnetic and hadronic sections, and extends
the coverage to $|\eta|<4.9$. 
The calorimeter system is surrounded by a muon spectrometer. 
Three layers of precision tracking chambers, consisting of drift tubes and cathode strip chambers, enable precise
muon track measurements in the pseudorapidity range of $|\eta| < 2.7$, and resistive-plate and thin-gap chambers provide
muon triggering capability in the range of $|\eta| < 2.4$.

The ATLAS trigger system uses three consecutive levels. The Level-1 triggers are hardware-based 
and use coarse detector information to identify regions of interest, whereas the Level-2 triggers 
are based on fast online data reconstruction algorithms. Finally, the Event Filter triggers use 
offline data reconstruction algorithms. This analysis uses a  trigger that requires a jet with 
transverse momentum $\pt > 100 \gev$ at Level-1. At the Event Filter level, the scalar sum of 
the $\pt$ of all jets with $\pt > 30 \gev$ and $|\eta|<3.2$ is required to 
be larger than either $350\,\gev$ or $400\,\gev$, depending on the data-taking period. 
These triggers are fully efficient for the offline event selection used in this analysis.

\section{Theoretical prediction}\label{sec:theory}
The cross-sections for $W$ or $Z$ bosons associated with jets are calculated at next-to-leading order (NLO) using the 
MCFM Monte Carlo (MC) program~\cite{Campbell:1999ah}. The calculation uses the CT10 parton 
distribution function (PDF) set~\cite{Lai:2010vv}.
The calculated $W$ production cross-section is approximately three times that of $Z$ production.
The total $W$ and $Z$ production cross-sections are then multiplied by the hadronic
$W$ and $Z$ branching fractions~\cite{Beringer:1900zz} to obtain the prediction for the hadronic 
\WplusZ cross-section for $\pT > 320~\gev$ and $|\eta|< 1.9$ of $\sigma_{W+Z}= 5.1\pm 0.5\,{\rm pb}$,
where the uncertainty of the calculation is described below. 

For the theoretical prediction the renormalization and factorization scales are dynamically  set to 
$H_{\rm T}/2$ of the event, where $H_{\rm T}=\sum_{i=1}^{n} p_{{\rm T},i}$ is defined as the scalar sum of the 
$p_{{\rm T},i}$  of the $n$ particles in the final state. 
The systematic uncertainty in the predicted cross-section due to higher-order corrections is estimated 
by independent variation of the renormalization and factorization scales between $0.5$ and $2.0$ times 
the nominal scale. Uncertainties in the prediction due to PDF uncertainties are computed from the 
52 CT10 eigenvectors at 68\% confidence level. The contribution due to the uncertainty in the value of 
the strong coupling constant \alphas\ is negligible. The total uncertainty in the theoretical prediction 
is computed by summing in quadrature the scale and the PDF uncertainties.

Although the calculation is performed at NLO, the prediction does not include contributions from 
radiative emissions of the quarks that originate from the W/Z boson decay or collinear $W$ emission inside quark and gluon jets.
The effect of a virtual photon is not included in the $Z$ cross-section.
The process $q\bar{q}\to \gamma^*\to q\bar{q}$ is estimated to constitute a negligible background.
The $W$ bosons from top-quark decays as well as the $W/Z$ bosons from diboson production are considered 
as background to the measurement.

\section{Data sample and Monte Carlo simulation}\label{sec:MC}
The data sample studied in this paper was recorded with the ATLAS detector at a \pp centre-of-mass energy 
of $7\,\tev$ in 2011,  and corresponds to an integrated luminosity of $4.6\pm 0.1$~fb$^{-1}$~\cite{Aad:2013ucp}.
Only data taken with all relevant detector sub-systems operational are used. 
Following basic data-quality checks, further event cleaning is performed by demanding that jets used in the 
analysis do not originate from instrumental effects, such as large noise signals in 
one or several channels of the hadronic end-cap calorimeter, or coherent noise in the electromagnetic calorimeter, 
or from non-collision background. Events are also required to have a reconstructed collision  vertex with at least three associated 
tracks, each with a transverse momentum greater than 400 MeV.

Simulated event samples of vector-boson production and of jets 
 with large transverse momentum produced via strong interactions  are used in the analysis. They are simulated utilizing different event 
generators, parton showering and hadronization models and various tunes of other soft model parameters, 
such as those of the underlying event, in order to compare to the features of the selected events.

The default simulated \WZ\ signal events are  generated using \HERWIG~\cite{Corcella:2000bw} 
interfaced to \JIMMY~\cite{Butterworth:1996zw}, using the modified MRST LO** PDF set~\cite{Sherstnev:2007nd,Sherstnev:2008dm}.
\HERWIG is based on a  leading-order (LO) perturbative QCD calculation.
The signal cross-sections are scaled by a 
$K$-factor of 1.25 to match the cross-section values predicted by an
NLO perturbative QCD calculation using MCFM~\cite{Campbell:1999ah}
with the CT10 PDF set~\cite{Lai:2010vv} as quoted in section~\ref{sec:theory}.
For cross-checks and the study of systematic uncertainties, additional 
signal samples are generated using 
\PYTHIAEIGHT~\cite{Sjostrand:2007gs}  and \PYTHIASIX~\cite{Sjostrand:2006za} with ATLAS Minimum Bias Tune 1 and 2B 
(AMBT1 and AMBT2B)~\cite{Aad:2010ac}. 
\PYTHIAEIGHT uses a LO  CTEQ6L1~\cite{Pumplin:2002vw} PDF set and 
\PYTHIASIX uses a modified MRST LO** PDF set. 
The decay modes of the vector bosons included in the signal MC simulations are
$W\to q\bar{q}^\prime$ and $Z\to q\bar{q}$, where $q, q^\prime = u, c, d, s$ 
or $b$. The effect of virtual photon production is not included in the simulation of $Z$ signal events.

The default QCD jet background events are generated with \PYTHIAEIGHT. Alternative background 
samples are also generated using \HERWIGPP~\cite{Bahr:2008pv}, \PYTHIASIX with AMBT1, AMBT2B 
and PERUGIA 2011~\cite{Skands:2010ak} tunes and \POWHEG~\cite{Alioli:2010xd,Alioli:2010xa}. \HERWIGPP uses a 
different hadronization model from \HERWIG and a modified MRST LO** PDF set. 
\POWHEG  is based on an NLO calculation that is interfaced to the \PYTHIASIX  showering routines;
the CT10 NLO PDF set  is used  for the matrix element calculation and the CTEQ6L1 PDF set is used  to generate the parton shower.

Top-quark pair events and single-top events in the $Wt$-channel are simulated with 
MC@NLO 4.03~\cite{Frixione:2002ik} interfaced to \HERWIG and \JIMMY   
and using the CT10 PDF set.  
In addition, an alternative $t\bar{t}$ MC sample is generated using POWHEG 1.0 (patch 4).
The top-quark pair production cross-section is 
scaled to match the calculated value from a next-to-next-to-leading-order calculation~\cite{Czakon:2013goa}.
Single-top events in $s$- and $t$-channels are simulated with \POWHEG. Simulation of diboson events 
($W^+W^-$, $W^\pm Z$ , $ZZ$, $W\gamma$ and $Z\gamma$ ) is performed with \HERWIG with the MRST LO** PDF. 

All simulation samples are generated with pileup by overlaying simulated minimum bias events on each generated signal
and background event. The number of overlaid events is simulated such that
the distribution of the average number of interactions per 
\pp bunch crossing in the simulation matches that observed in the data. This average varies with data-taking 
period and ranges typically between 4 and 16. The generated samples are processed through the 
GEANT4~\cite{Agostinelli:2002hh} simulation of the ATLAS detector \cite{Aad:2010ah} and the standard ATLAS 
reconstruction software.

\section{Jet reconstruction and selection}
\label{SEC:SELECTION}
Jets are reconstructed using the anti-$k_t$ algorithm~\cite{Cacciari:2008gp} with a jet radius parameter 
$R = 0.6$. Topological clusters~\cite{Aad:2010ad} of energy deposits in the calorimeters are 
used as input to the clustering algorithm. The topological clusters are calibrated to the hadronic 
energy scale \cite{Aad:2012vm,Aad:2014bia}. Selected jets are required to have transverse 
momentum $\pt > 320 \gev$, pseudorapidity $|\eta| < 1.9$  and reconstructed jet mass $50 \gev < m_{\rm jet} < 140 \gev$. 
Studies~\cite{Aad:2013gja} show that for a hadronically decaying \WZ boson with $\pt > 320 \gev$, the angular separation of their decaying products 
tends to less than $R=0.6$.
The jet mass is calculated from the energies and momenta of 
the jet constituents as $m_{\rm jet}=\sqrt{(\sum_i E_i)^2- |\sum_i \vec{p}_i|^2}$ where $E_i$ and 
$\vec{p}_i$ are the energy and three-momentum of the $i^{\rm th}$ constituent. At detector level, 
the jet constituents are the topological clusters that are considered massless.

In this measurement, hadronically decaying boosted $W$ and $Z$ bosons are identified using their reconstructed 
jet mass. The $W$ ($Z$) jet mass distribution peaks around the $W$ ($Z$) mass value, 
while the jet mass distribution from QCD jet events has a much broader spectrum. However, the QCD jet production 
cross-section is several orders of magnitude larger than the SM \WplusZ production cross-section. According 
to MC simulation, the data after the preselection consist almost entirely of QCD jets with a tiny
fraction of signal events, and the jet mass alone does not provide sufficient discriminating power to 
distinguish \WZ jets from the large QCD jet background. A jet substructure method~\cite{Chen:2011ah} 
based on cluster information evaluated in the jet centre-of-mass frame is used to suppress the 
QCD jet background while keeping most of the \WZ jets. The centre-of-mass frame (rest frame) of a jet 
is defined as the frame where the four-momentum of the jet is equal to 
$p_{\rm rest}\equiv (m_{\rm jet}, 0, 0, 0)$.

The  topology of a $W$ or $Z$ jet in its centre-of-mass frame is expected to be different from that 
of a typical QCD jet. In the rest frame of a hadronically decaying $W$ or $Z$ boson, the constituent 
particles in most of the cases look like a back-to-back dijet event. On the other hand, a QCD jet 
acquires its mass through gluon ($g$) radiation and $g \to q\bar{q}$ splitting. In this case, the constituent 
particle distribution in the jet rest frame does not correspond to a physical state with a well-defined 
mass. Three jet-shape variables, all calculated using the energy clusters of a jet in its centre-of-mass frame, 
are studied: thrust minor~\cite{Brandt:1964sa,Farhi:1977sg},  
sphericity~\cite{Bjorken:1969wi} and aplanarity~\cite{Bjorken:1969wi}.   
A value of thrust minor $T_{\rm min}=0.5$ corresponds to an isotropic distribution of energies while $T_{\rm min}=0$ indicates a highly collimated energy deposits. 
The sphericity is defined such that it is bounded to be between 0 and 1 and the most isotropic events have values close to 1.0. 
The aplanarity takes values between 0 and 0.5 and isotropic events yield values near 0.5.
The definitions of these shape variables are given in \ref{sec:app}.

\begin{figure*}[h!]
    \includegraphics[width=0.49\textwidth] {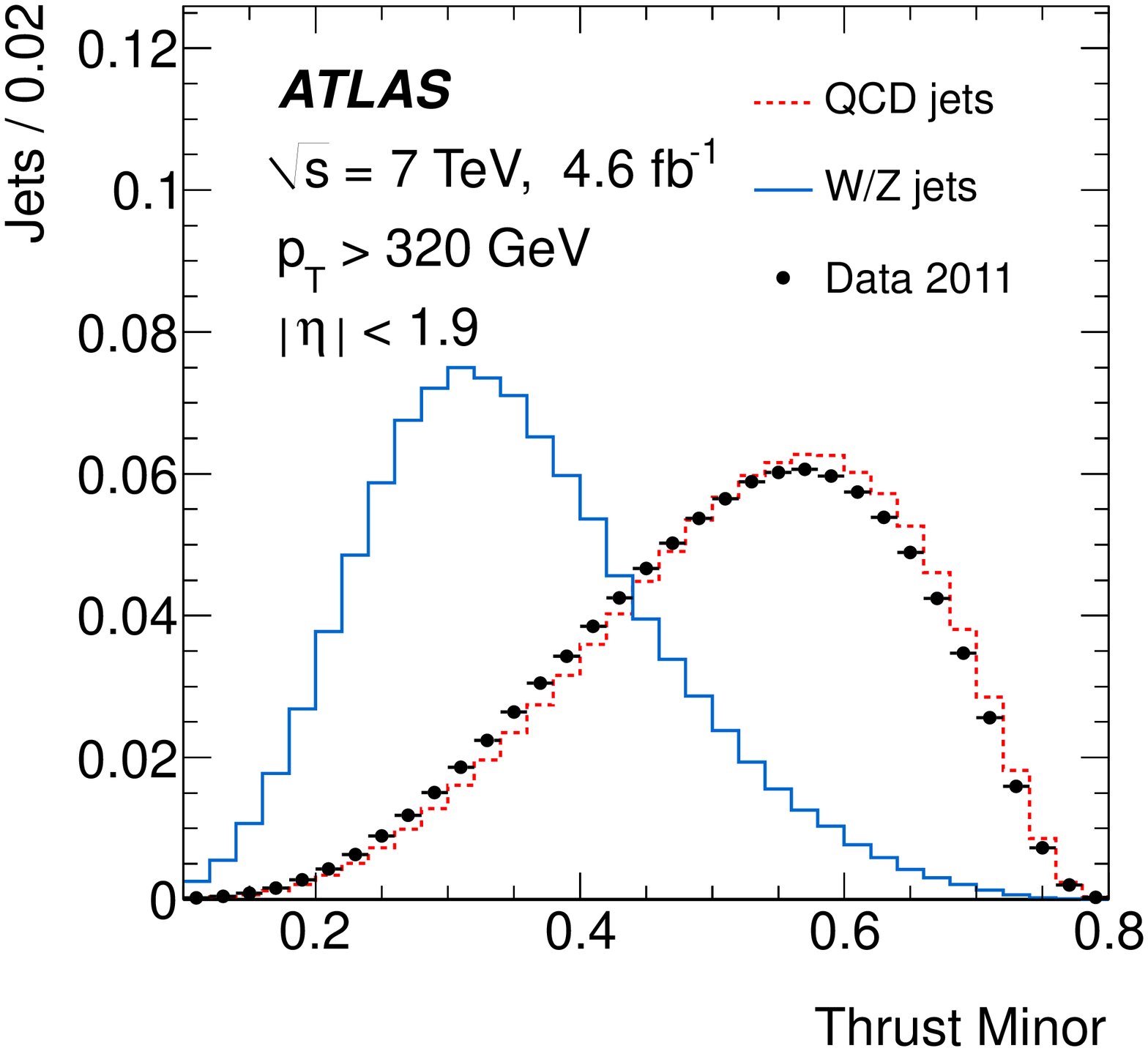}
    \includegraphics[width=0.49\textwidth] {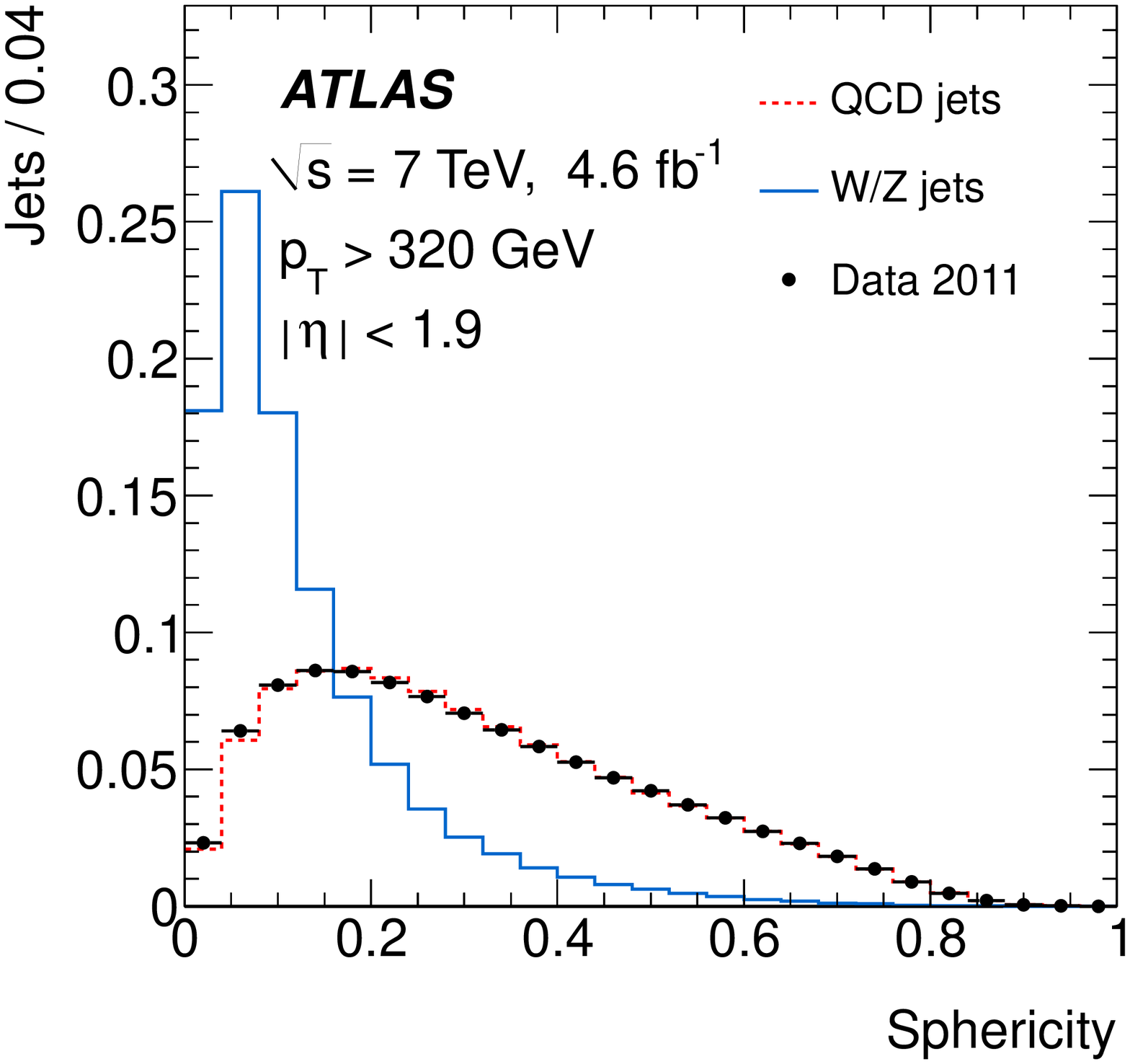}\\
    \includegraphics[width=0.49\textwidth] {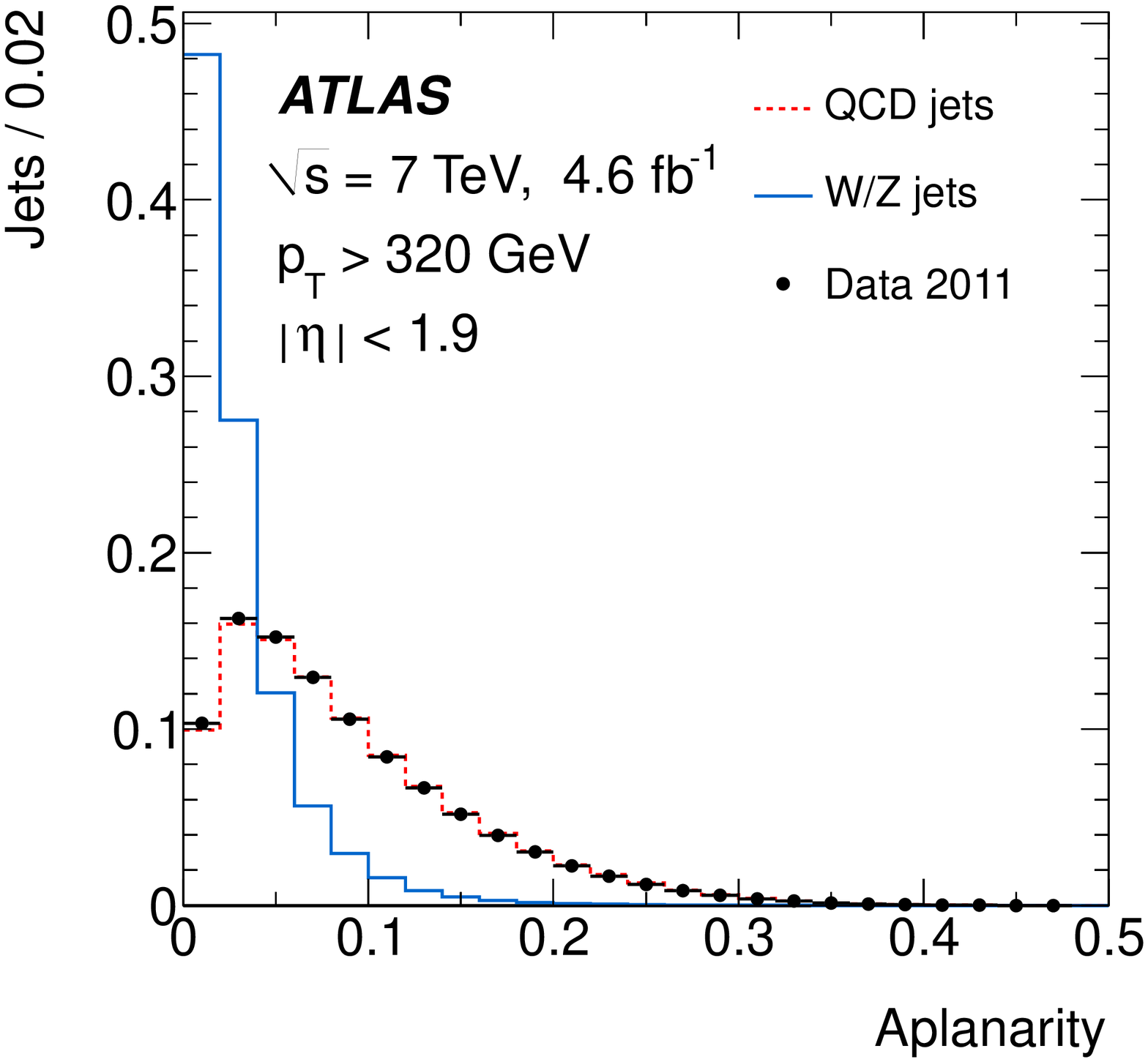}
    \includegraphics[width=0.49\textwidth] {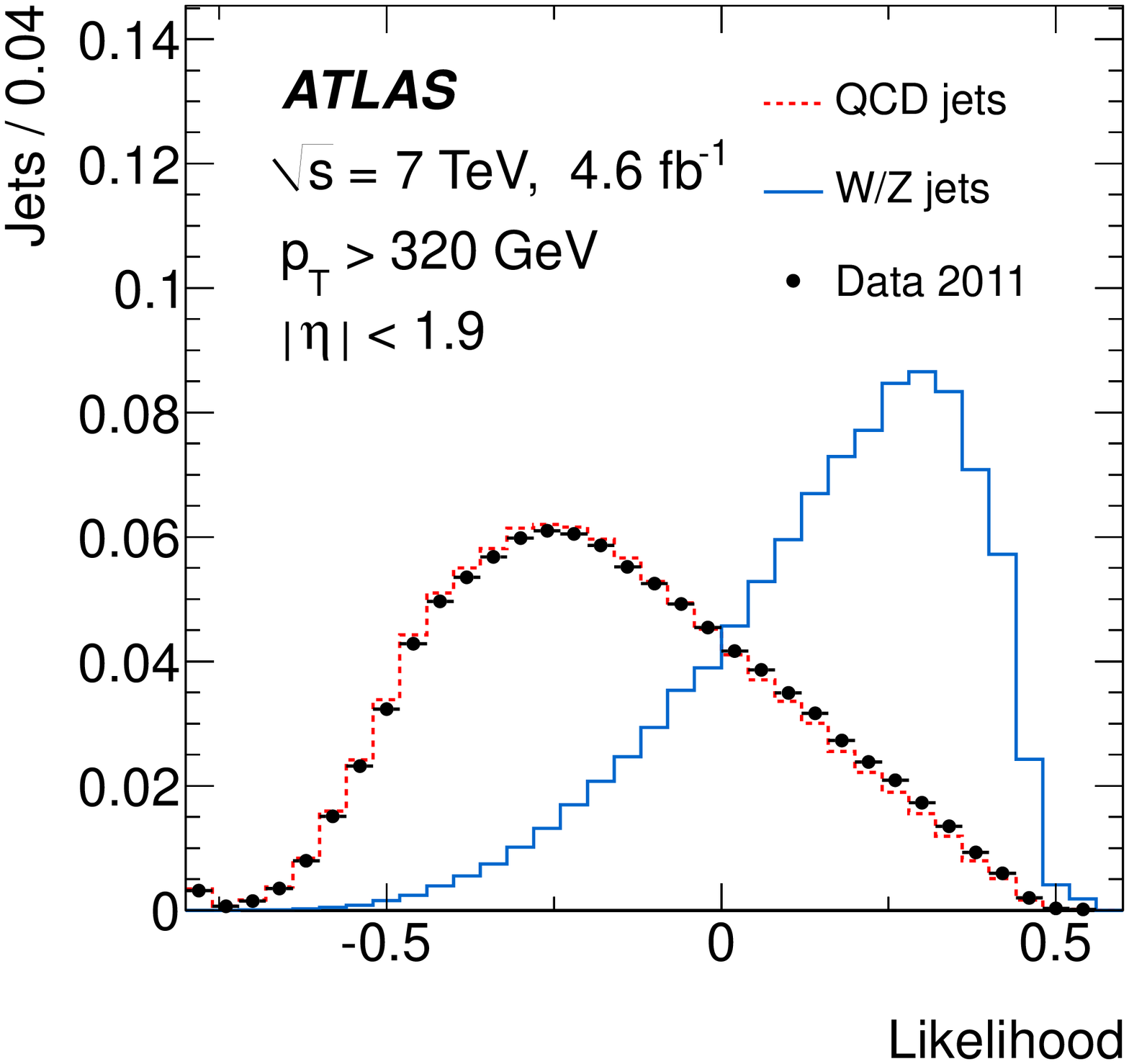}
    \caption{Unit area normalized distributions of the jet-shape variables and of the likelihood discriminant 
($L$) for \WZ jet signal (blue solid, produced using \HERWIG) and  QCD jet background (dashed red, produced using \PYTHIAEIGHT) 
in the MC samples and data (black dots).}
    \label{fig:jetShapesSignBckg} 
\end{figure*}

The distributions of the jet-shape variables  are shown in figure~\ref{fig:jetShapesSignBckg} for the 
\WZ jet signal,  QCD jet background and data. The distributions of the \WZ jet signal exhibit the 
characteristics of a back-to-back two-body topology, while those of the QCD jet background indicate 
a more isotropic distribution. The comparison of aplanarity and sphericity in data with 
simulated QCD jet events shows excellent agreement while there are small discrepancies between the 
thrust minor distributions. The comparison of data with different event generators and sets of 
generator parameters has also been performed. Although none of the MC generators and tunes studied 
show full agreement with the data for all the shape variables, the data are always within 
the variations of the distributions between different MC samples.	

To exploit the power of these variables to discriminate between \WZ jets and QCD jet 
background, a likelihood discriminant ($L$) is derived using the three jet-shape variables  as inputs, where the correlations between the variables
are ignored.
The likelihood ratio $L(i)$ for jet $i$ is defined by
{\par\nobreak\noindent}
\begin{equation}
L(i)= -\ln \frac{\mathcal L_s(i)}{\mathcal L_s(i)+\mathcal L_b(i)},
\label{likratio}
\end{equation}
with
\begin{equation}
\mathcal L_{s(b)}(i)=\prod_{k=1}^{3}p_{{\rm {s(b)}},k}(x_k(i)),
\end{equation}
where $p_{{\rm {s(b)}},k}$ is the normalized signal (background) probability density function (pdf) based on the default MC samples
for the $k^{\rm th}$ input variable $x_k$. The distributions of the likelihood discriminant are shown 
in figure~\ref{fig:jetShapesSignBckg} for \WZ jet signal,  QCD jet background and data. 
For the final event selection,  a cut on the 
likelihood discriminant is made: the optimal cut value is found by maximizing the statistical 
significance,  $S/\sqrt{S+B}$, where $S$ and $B$ are respectively the numbers of \WZ jet signal 
candidates and QCD jet background candidates predicted by the simulation. Candidate jets are 
required to have   $L>0.15$, which corresponds to 
$56\,\%$ signal efficiency and $89\,\%$ background rejection rate. 
In 2.5\% of the events in data, more than one jet candidate is found. After all the event selection 
criteria are applied, the final data sample consists of 590617 selected jets.

\section{\WplusZ cross-section measurement}\label{sec:xsec}

\subsection{Modelling of jet mass distributions}\label{sec:fit}

\begin{figure*}[t!]
\begin{center}
    \includegraphics[width=0.49\textwidth] {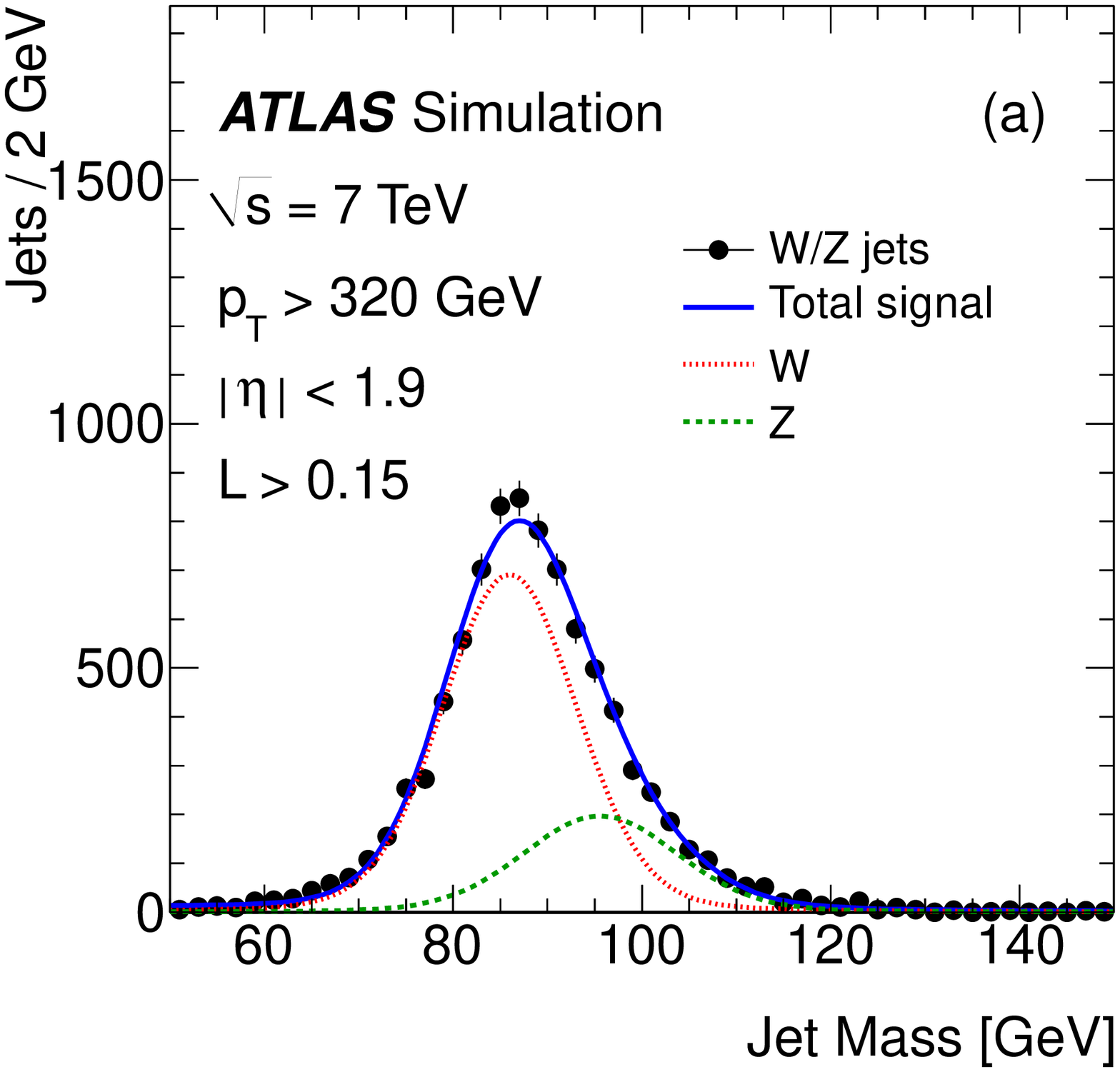}
    \includegraphics[width=0.49\textwidth] {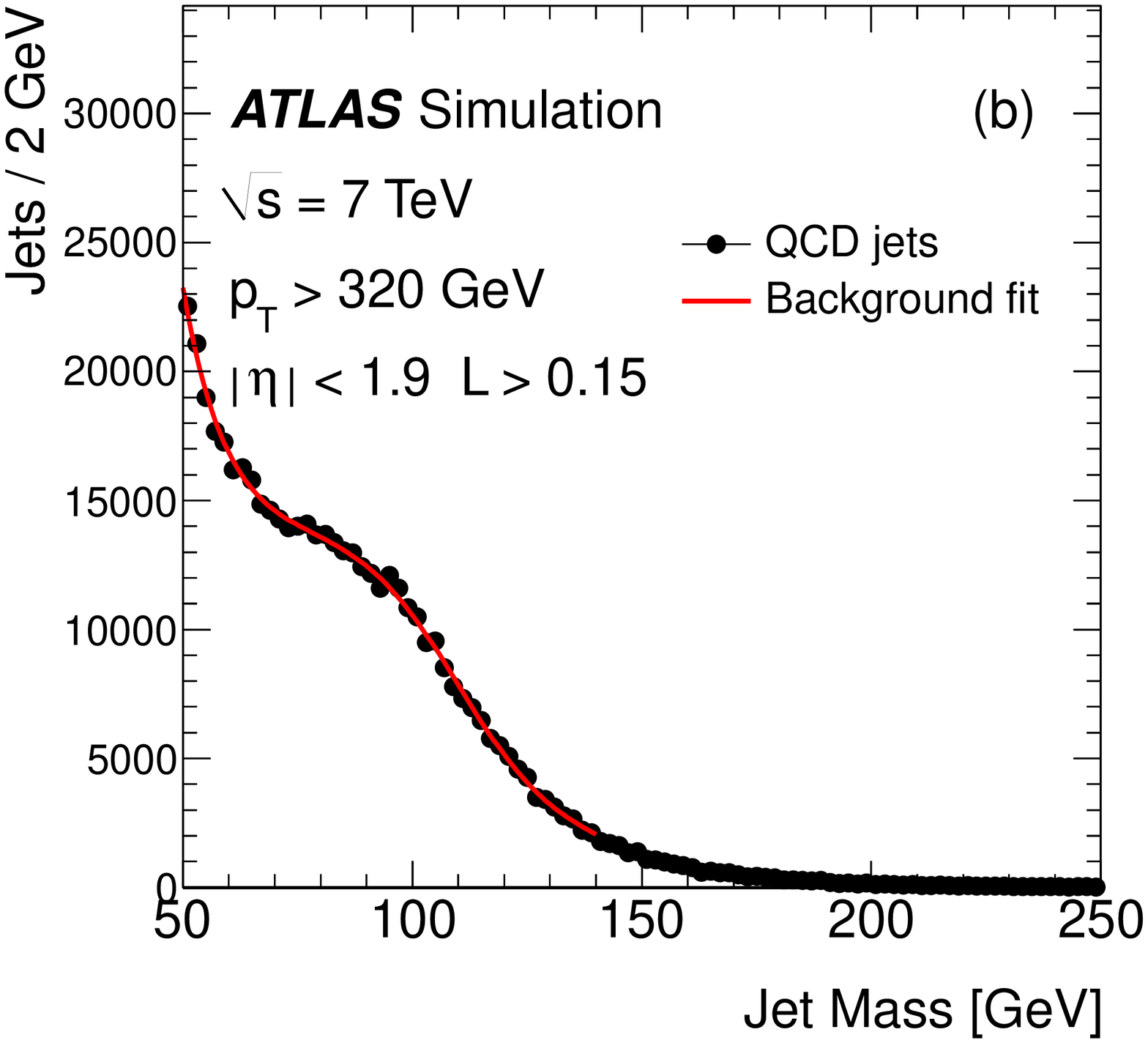}
    \includegraphics[width=0.49\textwidth] {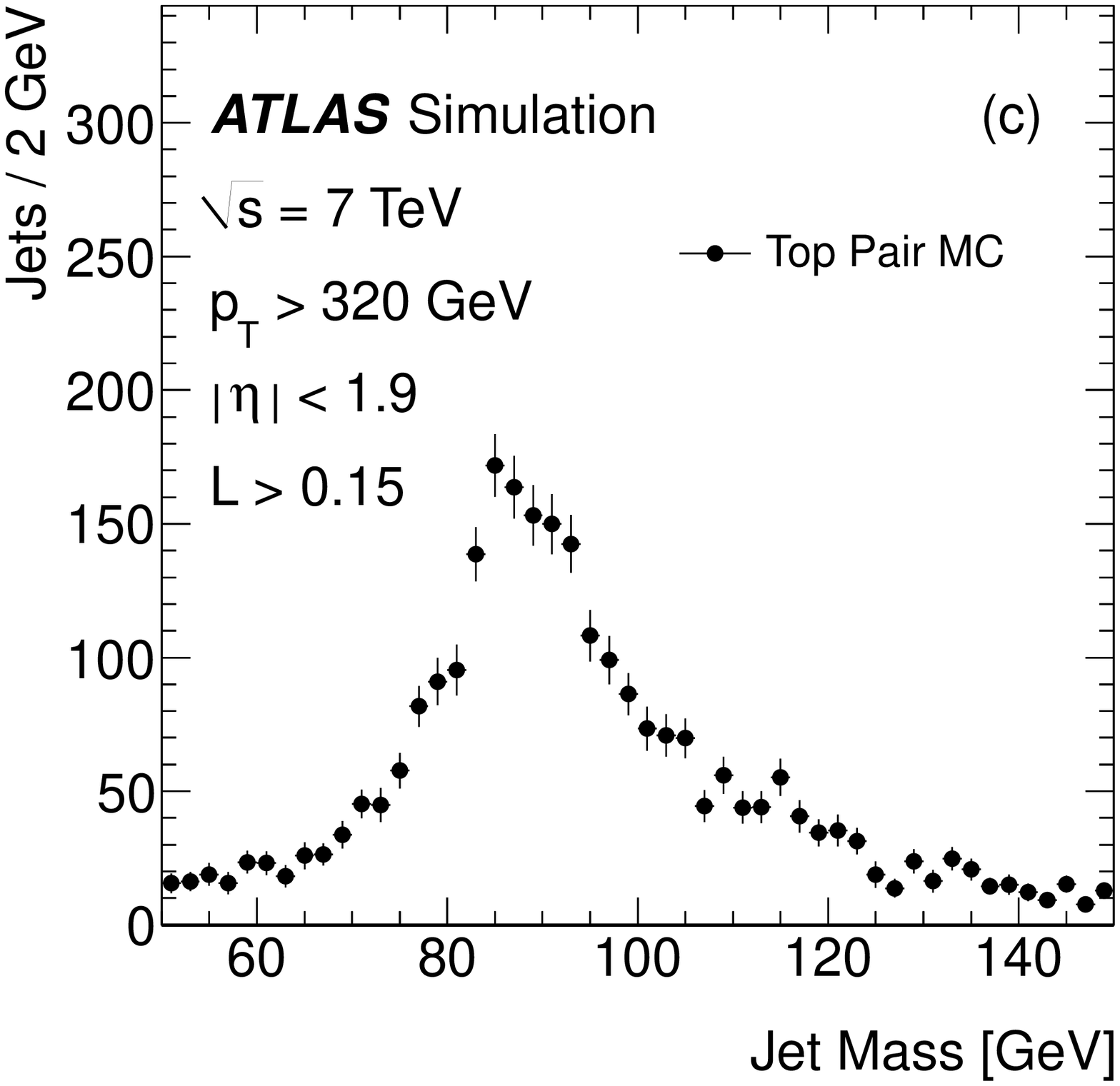}
 \caption{Jet mass distribution in simulation for (a) hadronically decaying $W$ and $Z$ bosons, 
(b) QCD jet background and (c) hadronically decaying $W$ bosons from top-quark pair events. The QCD jet
background and \WZ signal distributions are fitted with the pdfs described in the text.
The fit results are shown as solid lines. For the signal, the contributions from $W$ (dotted line) 
and $Z$ (dashed line) jets are shown. The distributions are normalized to the number of events 
expected in the dataset used and the uncertainties are statistical only.}
    \label{fig:pdfs}
\end{center}
\end{figure*}

The \WZ jet signal yield is extracted using a binned maximum likelihood fit to the jet mass 
distribution of the selected jets in the data. 
The probability density functions for the \WZ  signal and the background are modelled as follows.

The $W$ jet and $Z$ jet signal pdfs are each modelled as a Breit--Wigner function convolved with
a Gaussian function in order to take into account detector resolution effects.
Due to additional contributions from pileup, underlying event and hadronization, the peak 
positions of the reconstructed $W$ and $Z$ signal jets are higher than the masses of the 
$W$ and $Z$ bosons. The parameters of the signal pdfs: the peak positions, 
widths and relative fractions of the $W$ and $Z$ rates, are obtained from 
a fit to the selected \WZ jets in the simulated events. In the fit to the data, the only free parameter 
affecting the signal is the combined total rate of $W$ and $Z$ bosons. 
In figure~\ref{fig:pdfs}(a) the jet mass distribution for \WZ jets in simulation overlaid with the signal pdfs is shown. 

The dominant background component in the jet mass distribution comes from the QCD jets.
The QCD jet mass distributions in the default MC sample and the alternative 
samples using different MC generators and tunes described in section~\ref{sec:MC} are similar 
and can all be described by the same analytic function. The same function is used to describe the data. 
However, the values of the parameters of the function 
differ slightly between the various MC samples and can also be different for the data,
and therefore left free in the fit to the data.
The jet mass distributions are parameterized by the sum of two exponential decay functions and a sigmoid 
function:
$S(\bar{m}) = \bar{m} / \sqrt{1+\bar {m}^2},$
where  $\bar{m} = {(m_{\rm jet}-m_0)}/{\sigma_m}$; 
 the parameters $m_0$ and $\sigma_m$  represent the position of the 
inflection point of the sigmoid function and the slope at the inflection point respectively. 
In Figure \ref{fig:pdfs}(b) the jet mass distribution for simulated QCD jet background 
and the fit result obtained using the background pdf are shown. 
The jet mass distribution of the QCD jet background displays a shoulder structure that is
described by the sigmoid function. 
This feature is related to the \pT\  and $L$ requirements, the kinematics and internal structure 
 of the selected jets, and to the distance parameter of the jet reconstruction algorithm.
 The variations of the shoulder structure observed in data with respect to different kinematic selection requirements 
 and the distance parameter of the jet reconstruction algorithm are well reproduced in the MC simulation of QCD jet production.
 
To determine the direct \WplusZ production cross-section, background from top-quark decays to $W$ bosons must be subtracted. 
The top-quark pair ($t\bar{t}$)  component is modelled using a one-dimensional histogram based on the 
simulation, as shown in figure~\ref{fig:pdfs}(c). 
After all the event selection criteria are applied, 2700 jets are predicted from the $t\bar{t}$ MC sample 
in the range of $50\gev < m_{\rm jet} < 140\gev$. About half of these jets populate a peak near the signal region with 
some additional enhancement at higher jet mass, which is due to partial overlap of the $W$ jet with 
a nearby $b$-jet. Since the peaking component of this background  is small comparing to the expected signal yield, 
its yield and shape are fixed to the ones predicted by the simulation.

The contributions from other background sources such as single-top production and diboson ($WW$, $WZ$, $ZZ$, 
$W\gamma$ and $Z\gamma$) production are expected to be very small according to the simulation, but they 
also produce peaks similar to the signal. The simulation predicts $190$ and $180$ \WZ candidates from 
single-top production and diboson production respectively. These background components are not explicitly 
considered in the fit. Instead, their expected contributions are subtracted from the fitted signal yield. 

\subsection{Fit to the \WZ jet mass distribution}
\begin{figure}[t!]
\begin{center}
    \includegraphics[width=.99\columnwidth] {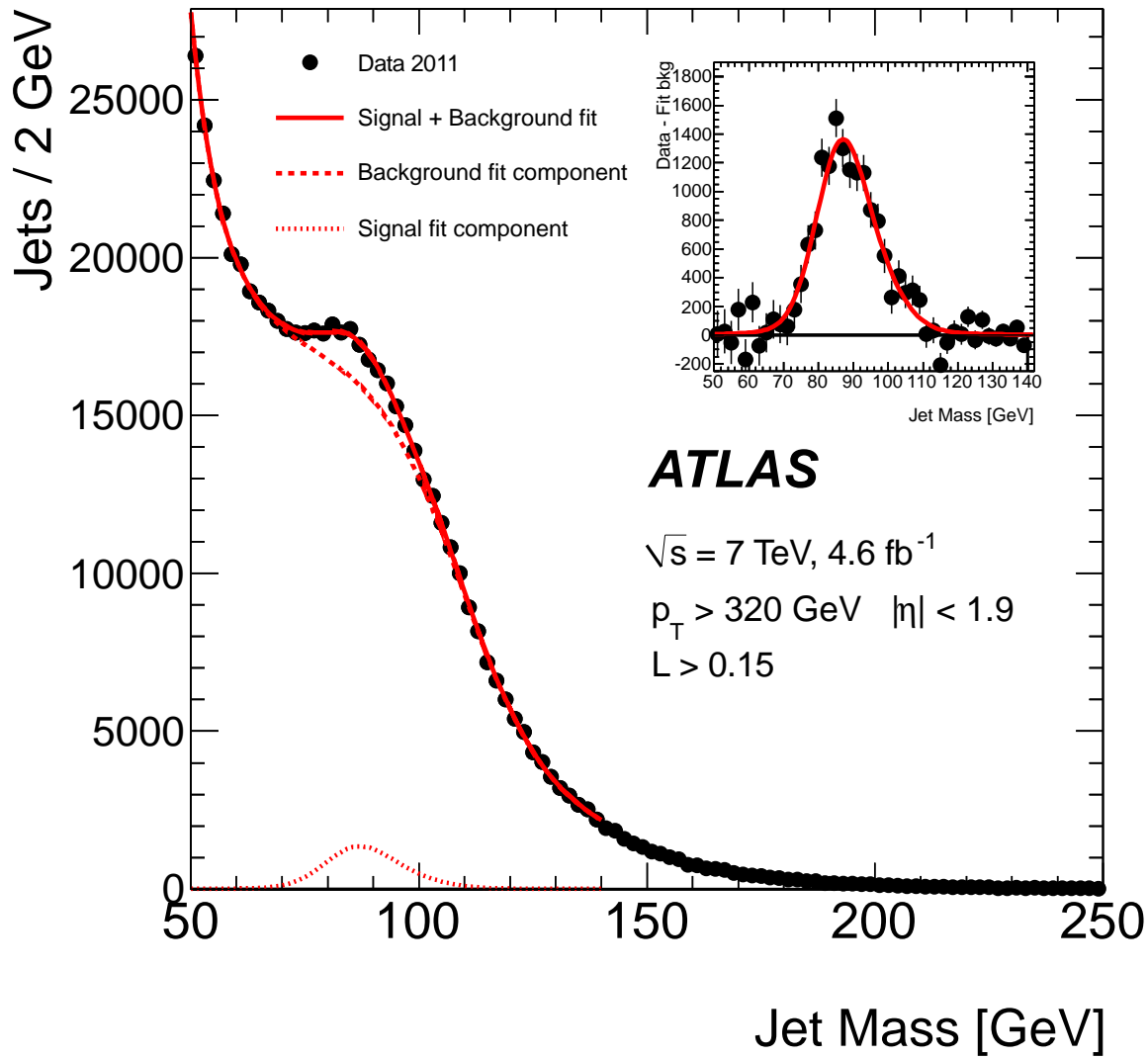}
 \caption{Jet mass distribution of the selected \WZ jets overlaid with the fit result
(for illustration, the jets with $m_{\rm jet} >140\, \gev$ are also shown).
The fit range is limited to $ 50\,\gev <  m_{\rm jet} < 140\,\gev$.
The background pdf component (dashed line), the signal pdf component (dotted line) and the 
total pdf (solid line) are shown. The data minus the fitted background component is shown in the 
inset.}
    \label{fig:fitResult}
\end{center}
\end{figure}

The observed jet mass distribution is fitted to the sum of \WZ signal and background pdfs. 
In addition to the combined \WZ signal yield, all the parameters of the QCD background
pdf are allowed to  float in the fit. They are: $m_0$ and $\sigma_m$ of the sigmoid function $S(\bar{m})$;
the slope parameters of the two exponential functions and the relative fractions of the QCD background 
components. 
The fit result is shown in figure \ref{fig:fitResult}. The $\chi^2$ per degree of freedom of the fit 
is  $\chi^2/\rm ndf=41.4/38$, which corresponds to a $\chi^2$-probability of 32\%. The total number of \WZ jet signal events 
in the range $50 \gev < m_{\rm jet} < 140 \gev$ obtained from the fit, after subtracting the diboson 
and single-top contributions, is $N^{W+Z}= 14200 \pm 1300$, where the uncertainty is the statistical 
uncertainty in the fitted signal yield.

The cross-section is calculated using 
{\par\nobreak\noindent}
\begin{equation}
\sigma_{W+Z}=N^{W+Z}/(\mathcal L\cdot \varepsilon),
\end{equation}
where $\mathcal L$ is the integrated luminosity. The efficiency $\varepsilon$ is estimated from simulation 
using the HERWIG generator and is defined as $\varepsilon={N_{\rm reco}^{W+Z}}/{ N_{\rm gen}^{W+Z}}$ 
where $N_{\rm reco}^{W+Z}$ is the number of $W$ and $Z$ jets in simulation passing the selection 
cuts and $N_{\rm gen}^{W+Z}$ is the number of  generated $W$ and $Z$ bosons with transverse momentum 
$p_{\rm T}>320\,\gev$   and  pseudorapidity $|\eta|<1.9$ at the generator level. 
The efficiency is estimated to be $0.36\pm 0.02$, where the uncertainty is due to the jet energy scale, 
jet energy resolution, and the variation between the efficiencies provided by the different MC generators 
and settings of generator parameters; these uncertainties are discussed later. 
The sum of the $W$ and $Z$ hadronic cross-sections is measured to be
{\par\nobreak\noindent}
\[\sigma_{W+Z}= 8.5 \pm 0.8\stat ~{\rm pb},\]
for $W$ and $Z$ bosons with $p_{\rm T}>320\,\gev$ and $|\eta|<1.9$,
where the uncertainty here is statistical only.

\subsection{Systematic uncertainties}\label{sec:syst}
The systematic uncertainty in the measured cross-section has contributions from the various 
sources listed in table \ref{tab:syst}.

\begin{table}[!b]
\begin{center}
\begin{tabular}{lr}
\toprule
Sources & $\sigma_{W+Z} $  \\
\midrule
MC modelling                & $4.4\,\%$ \\
Background pdf             & $8.8\,\%$ \\
Signal pdf                 & $5\,\%$\\
Jet energy scale           & $3.7\,\%$ \\
Jet energy resolution      & $<1\,\%$ \\
Jet mass scale             & $2.2\,\%$  \\
Jet mass resolution        & $12.6\,\%$  \\
$t\bar{t}$ contribution    & $2.8\,\%$   \\
Single-top and diboson contribution & $<1\,\%$   \\
$W$ and $Z$ relative yield & $2.9\,\%$ \\
Luminosity                 & $1.8\,\%$ \\
\midrule
Total &                    $18\,\%$    \\
\bottomrule
\end{tabular}\caption{Summary of the relative systematic uncertainties 
in the \WplusZ jet cross-section from different sources.}
\label{tab:syst}
\end{center}
\end{table}
The uncertainty in the selection efficiency due to the choice of generator and setting of 
generator parameters for the simulation is estimated by using the alternative MC samples
described in section~\ref{sec:MC}. The RMS spread in the efficiencies obtained from various 
generators and configurations with respect to the default ones is taken as the uncertainty.

In order to estimate the systematic uncertainty due to the choice of QCD background pdf,  
the fit to the data is repeated with different background models that include: adding an 
exponential term to the default background pdf; removing one of the two exponential terms 
from the default background pdf; replacing $S(\bar{m})$ in the default background pdf with 
a different sigmoid function, such as the complementary error function $\erf(\bar{m})$, the 
hyperbolic tangent $\tanh(\bar{m})$, or the arctangent $\arctan (\bar{m})$. 
All the fits including alternative background models describe the data
reasonably well and have a $\chi^2$-probability larger than 
$1\%$. The largest deviation of the fitted signal yield using different background pdfs with 
respect to the nominal fit is taken as the corresponding systematic uncertainty.

The uncertainty in the fitted signal yield due to the choice of the signal pdf is obtained by repeating 
the fit using one-dimensional histograms based on alternative MC simulations  to model the signal pdf. The 
largest deviation of the signal yield with respect to the nominal fit is assigned as systematic 
uncertainty.

The robustness of the fit has been studied with ensembles of pseudo-datasets composed of background and 
signal events obtained from the MC simulation. The number of background events is set to the value 
predicted by the simulation. The number of signal events is varied in the ensembles between zero 
and the signal yield observed in data. Fits to each of the pseudo-datasets with the default signal and 
background model are performed. No bias in the fitted signal yields with respect to the number of 
signal events present in the pseudo-datasets is observed.

Uncertainties in the simulation of the detector response are taken into account using dedicated studies of 
the reconstructed jets. The uncertainties considered are associated with 
(a) the jet energy scale,  
(b) the jet energy resolution, 
(c) the jet mass scale and 
(d) the jet mass resolution. 
The uncertainties in the measured cross-section due to (a) the jet energy scale~\cite{Aad:2011he}
and (b) the jet energy resolution~\cite{Aad:2012ag} are evaluated by computing the signal 
selection efficiency when varying the corresponding parameters within their uncertainties.
The uncertainty in the measured cross-section due to (c) the jet mass scale
is obtained from data through the introduction of a common offset $\Delta m$ to  
the $W$ and $Z$ signal pdf models (offset to the means of the Breit--Wigner functions) 
as a free parameter in the fit. 
The fitted value of the offset is $\Delta m= -0.45\pm 0.86 \,\gev$, compatible with zero
within the statistical uncertainty. The fit to the data is repeated with the value of
$\Delta m$ fixed to $-1.31\,\gev$ or $0.41\,\gev$. The larger deviation of the fitted
 signal yield with respect to the nominal fit is taken as the corresponding systematic uncertainty.
This estimate of the systematic uncertainty  has been 
cross-checked with a different technique using jets composed from tracks geometrically matched to 
calorimeter jets~\cite{ATLAS:2012am}.
The uncertainty in the measured cross-section due to (d) the jet mass resolution uncertainty is
 obtained by studying the jet mass resolution in  simulation. The generators and setting of generator 
parameters described in section~\ref{sec:MC} are used to study the effect of the parton shower and 
the hadronization 
model on the mass resolution. Instrumental effects are considered using a simulation of the ATLAS 
detector with a different amount of passive material. The detector response is  also studied in a 
simulation where a different  modelling of interactions of high-energy hadrons is 
implemented~\cite{Andersson:1992iq,Heikkinen:2003sc,Amelin:1989wa}. 
The total systematic uncertainty due to the jet mass resolution is obtained by adding in quadrature 
the uncertainties from the above-mentioned sources, and the fit to the data is repeated with an increased 
and decreased value of the peak resolution. The larger deviation of the fitted signal yield with respect 
to the nominal fit is taken as the corresponding systematic uncertainty.

Uncertainties in the fitted signal yield due to the $t\bar{t}$ contribution are assessed by
changing the expected $t\bar{t}$ contribution within the theoretical uncertainty in the inclusive
top-quark pair production cross-section~\cite{Czakon:2013goa} in the fit, and by repeating the fit using alternative 
$t\bar{t}$ MC samples generated with different initial/final state radiation settings or with a different 
algorithm (\POWHEG). The deviations in the fitted signal yield with respect to the nominal fit 
result are added in quadrature and taken as the corresponding systematic uncertainty.
Similarly, the small expected single-top and diboson yields are varied by $\pm 50\%$. The shifts in the 
signal yield are assigned as systematic uncertainties.

Theoretical uncertainties in the fitted signal yield arise from fixing the ratio of the $W$ to $Z$ 
event yield in the fit. The fit is repeated varying the relative signal yield of $W$ and $Z$ bosons 
within its theoretical uncertainty of 2\%~\cite{Campbell:1999ah} and the fitted yield variation with 
respect to the nominal fit result  is assigned as a systematic uncertainty. 

The uncertainty in the luminosity is $1.8\%$~\cite{Aad:2013ucp}. 
Other systematic sources considered in the measurement include the finite size of the MC sample, and pileup effects. 
All of them are found to have negligible effects on the measurement ($<1\%$).
The total systematic uncertainty in  the cross-section measurement  is calculated to be $18\%$ by adding all the systematic uncertainties 
in quadrature. The total systematic uncertainty is dominated by the  uncertainty in the  jet mass  resolution.

\subsection{\WplusZ cross-section result}\label{sec:finalresult}
The sum of the cross-sections of $W$  and $Z$ bosons decaying hadronically is measured to be
\[\sigma_{W+Z}= 8.5 \pm 0.8 \stat \pm 1.5 \syst~{\rm pb},\]
for $W$ and $Z$ bosons with $p_{\rm T}>320\,\gev$ and $|\eta|<1.9$. 

The measured cross-section is found to be in agreement with the theoretical prediction, based 
on an NLO MCFM calculation for $W/Z$ production in association with jets, of 
$\sigma_{W+Z}= 5.1 \pm 0.5\,{\rm pb}$ within 2 standard deviations. The 
uncertainty in the theoretical cross-section represents missing higher-order contributions estimated 
by varying the factorization and renormalization scales and uncertainties in the PDF of 
the proton, as detailed in section~\ref{sec:theory}.

\section{Study of the effects of various jet grooming techniques}

\def\LHgroom{\ensuremath{L_{\text{groom}}}\xspace}
\def\LHprune{\ensuremath{L_{\text{pruned}}}\xspace}
\def\LHtrim{\ensuremath{L_{\text{trimmed}}}\xspace}

The event sample, selected as described in section~\ref{SEC:SELECTION}, constitutes a 
sample of jets containing a relatively high fraction of boosted $W$ and $Z$ bosons decaying 
hadronically.  It is interesting to use such a sample to study the performance of various proposed jet grooming 
techniques.  The grooming techniques studied here are 
pruning \cite{Ellis:2009me} and trimming \cite{Krohn:2009th}, designed to suppress soft QCD 
radiation in jets, and area subtraction \cite{areasub} designed to correct for the effects of 
pileup.  The grooming techniques and their implementation are described in detail in~\ref{sec:appC}.

\subsection{Methodology}
The grooming algorithms are applied to jets that pass the selection for the 
cross-section measurement except for the likelihood ratio requirement. 
The pruning and trimming algorithms reduce the jet constituents used in the calculation of 
the likelihood ratio and jet mass. For these grooming algorithms, likelihood ratios (\LHprune and \LHtrim)
are calculated after grooming. The cut values after pruning or trimming ($\LHprune >0.16$ and 
$\LHtrim > 0.16$) are chosen to obtain the same background rejection (89\%) as the cut on 
$L$ for the ungroomed jets. 
The cuts on \LHprune and \LHtrim are applied to the pruned and trimmed jets respectively, and 
the jet mass distributions are studied. For the area subtraction algorithm, the default jet selection 
including the $L$ requirement is used, but the jet mass is recalculated after the estimated 
pileup contribution is subtracted from the jet. No attempt is made to optimize the 
analysis for the grooming techniques studied.

\subsection{Jet mass distributions}
\begin{figure}[h]
\begin{tabular}{cc}
\includegraphics[width=0.49\textwidth] {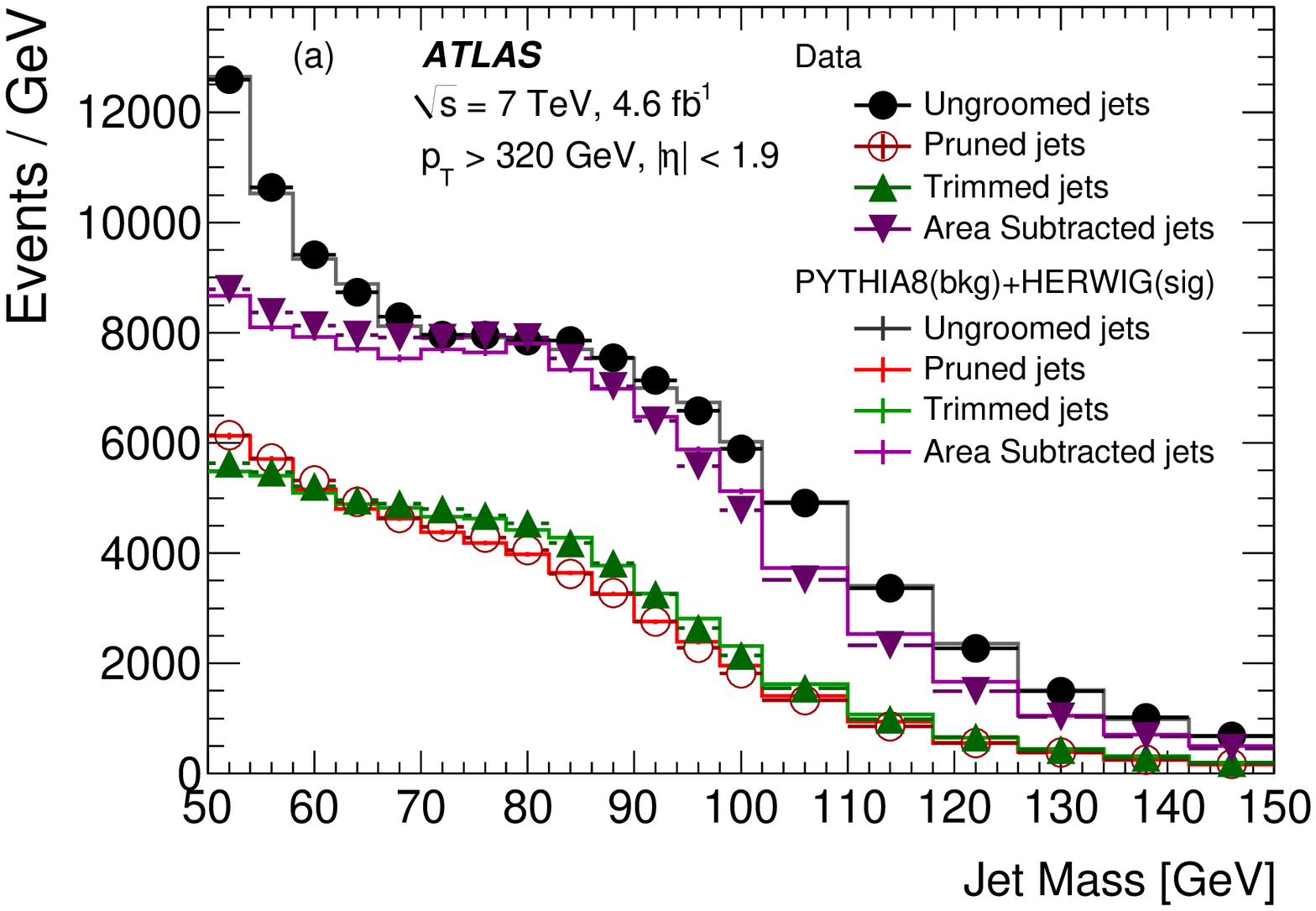} 
\includegraphics[width=0.49\textwidth] {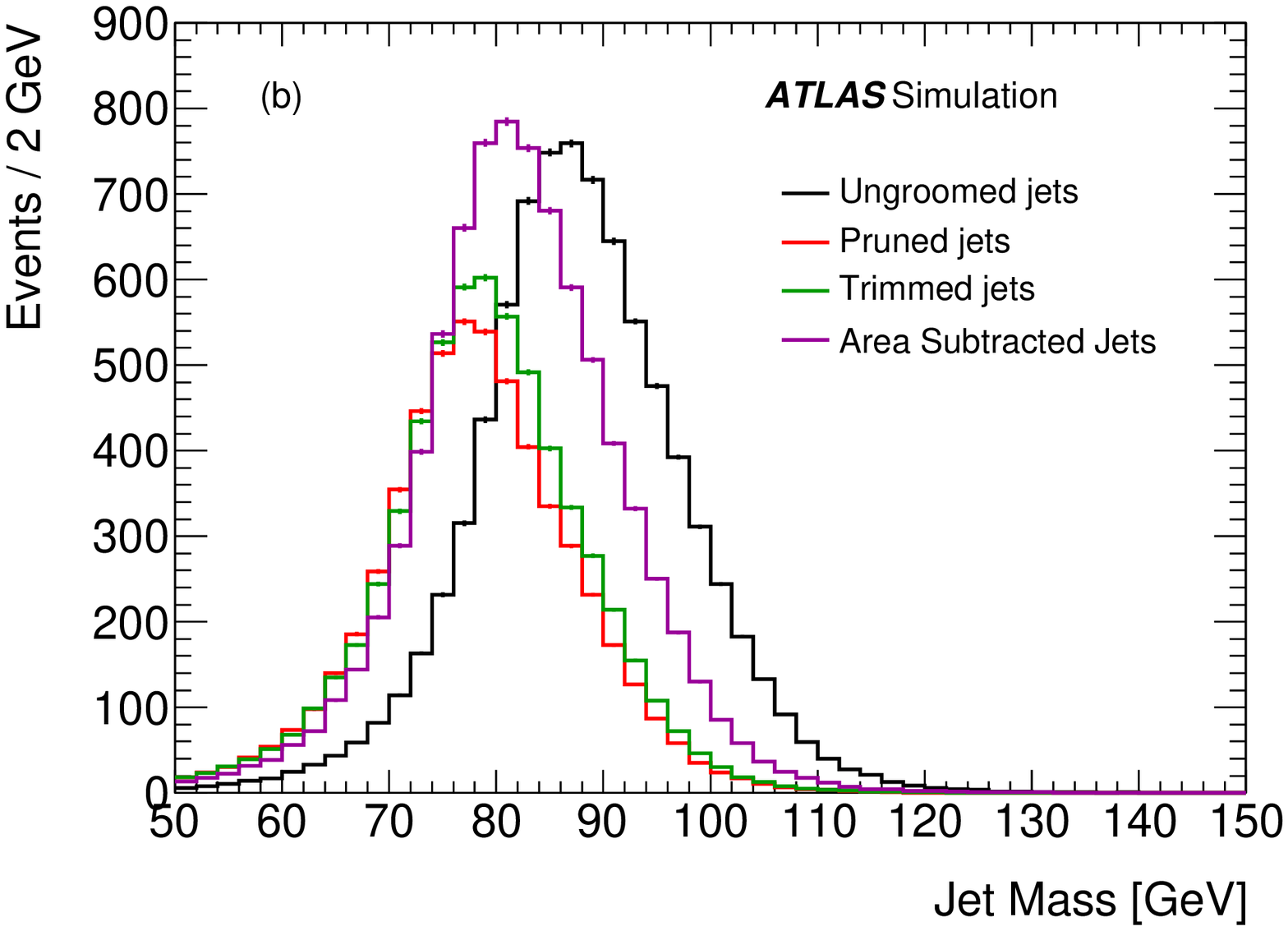} 
\end{tabular}
 \caption{The jet mass distributions (a) in the MC samples (\PYTHIAEIGHT for background plus \HERWIG for signal) and data and (b) for signal only, after cuts on the 
likelihood ratios calculated from the ungroomed, pruned, trimmed and area subtracted jet-shape 
variables are applied.}
 \label{fig:massgroomedlik}
\end{figure}

The jet mass distributions obtained after cutting on the new likelihood ratios (for trimming
and pruning) or pileup subtraction (for area subtraction) are shown in figure~\ref{fig:massgroomedlik}(a). 
Jet grooming causes an average shift to lower jet masses. A reduction in the number of selected 
jets with masses above 50~\gev\ by about 50\% compared to the ungroomed 
case is observed after trimming and pruning and by 15\% after area subtraction. The shoulder 
structure in the mass distribution of the QCD jet background is still present after jet grooming. 

Figure~\ref{fig:massgroomedlik}(b) shows the mass distribution for simulated \WZ jet signal events
after jet grooming. For area subtraction, the mean of the jet mass distribution is shifted 
lower by about 6~\gev\ and the number of selected jets is unchanged. After trimming and pruning, 
the mean is shifted lower by 8~\gev\ and 9~\gev\ respectively, and the number of selected jets 
is reduced by about 30\%. The ratio of the width to the mean of the \WZ jet mass distribution 
does not change significantly with grooming.

The shapes of the mass distributions of the selected jets in data are compared with simulation 
in figure~\ref{fig:massgroomedlik}(a). For the simulated jets, the 
predicted mass distributions of the \WZ jets and the QCD jets are added. 
The normalization of the \WZ jets sample is kept fixed to the NLO prediction described in 
section~\ref{sec:theory}, while the normalization of the QCD jets sample is adjusted such that 
the total number of selected jets in the simulation and the data agree.
The normalized \PYTHIAEIGHT QCD jet Monte Carlo and \HERWIG \WZ signal Monte Carlo provide 
reasonable descriptions of the groomed mass distributions while the ungroomed distribution 
is well described.

The statistical significance of the \WplusZ signal (assuming the theoretically predicted cross-section) 
remains about the same before and after grooming. For trimming and pruning, 
the reduction of background is offset by a loss in signal efficiency, 
while for area subtraction, the jet mass 
distributions remain similar, in the signal region, to the ungroomed sample. While systematic 
uncertainties have not been evaluated, the \WplusZ cross-section determined from the groomed 
jet samples, using the technique described in section~\ref{sec:xsec}, gives results compatible 
with that given in section~\ref{sec:finalresult} within statistical uncertainties.

\subsection{Pileup dependence}
\begin{figure}[!ht]
\begin{center}
\includegraphics[width=0.45\textwidth] {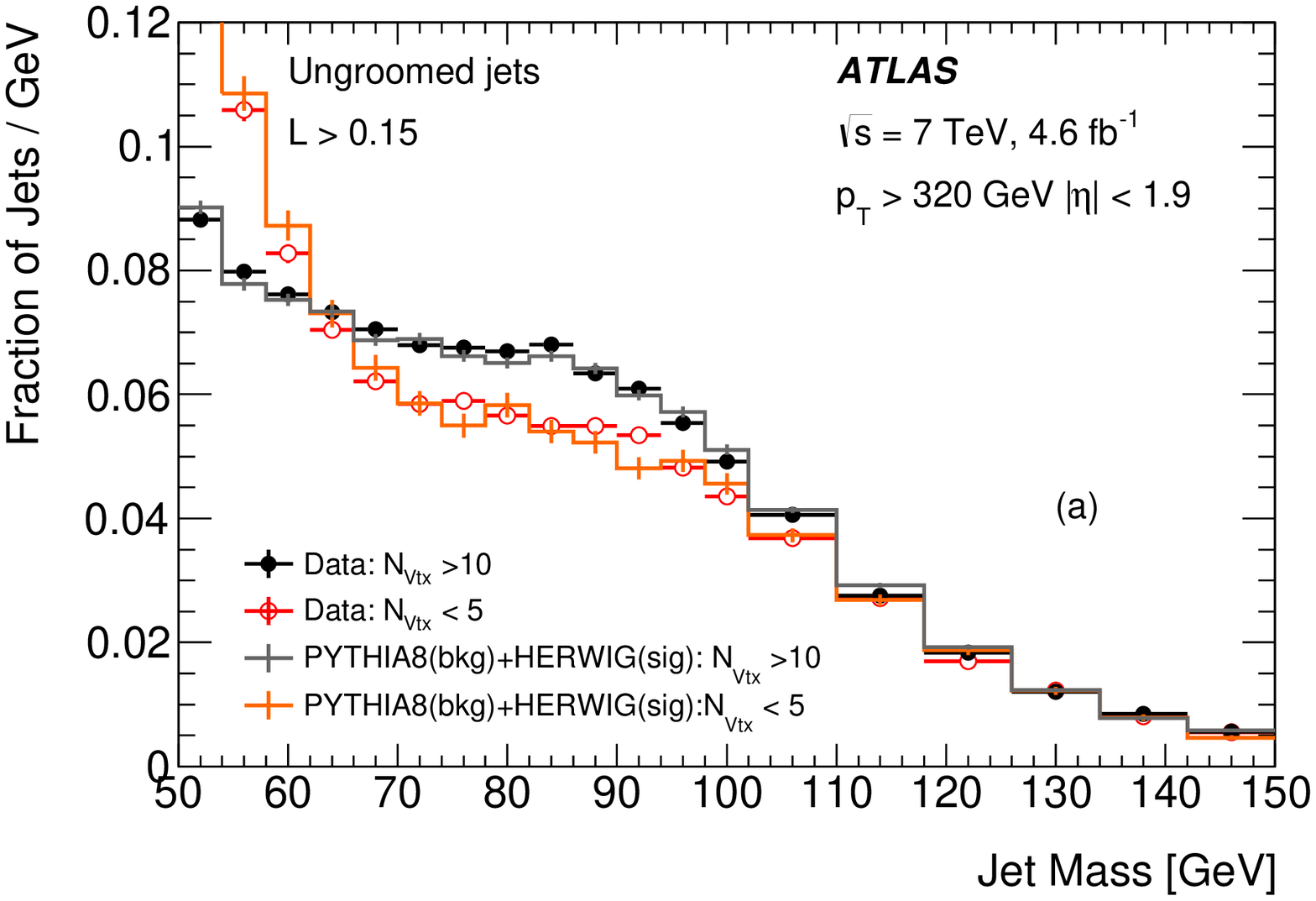} 
\includegraphics[width=0.45\textwidth] {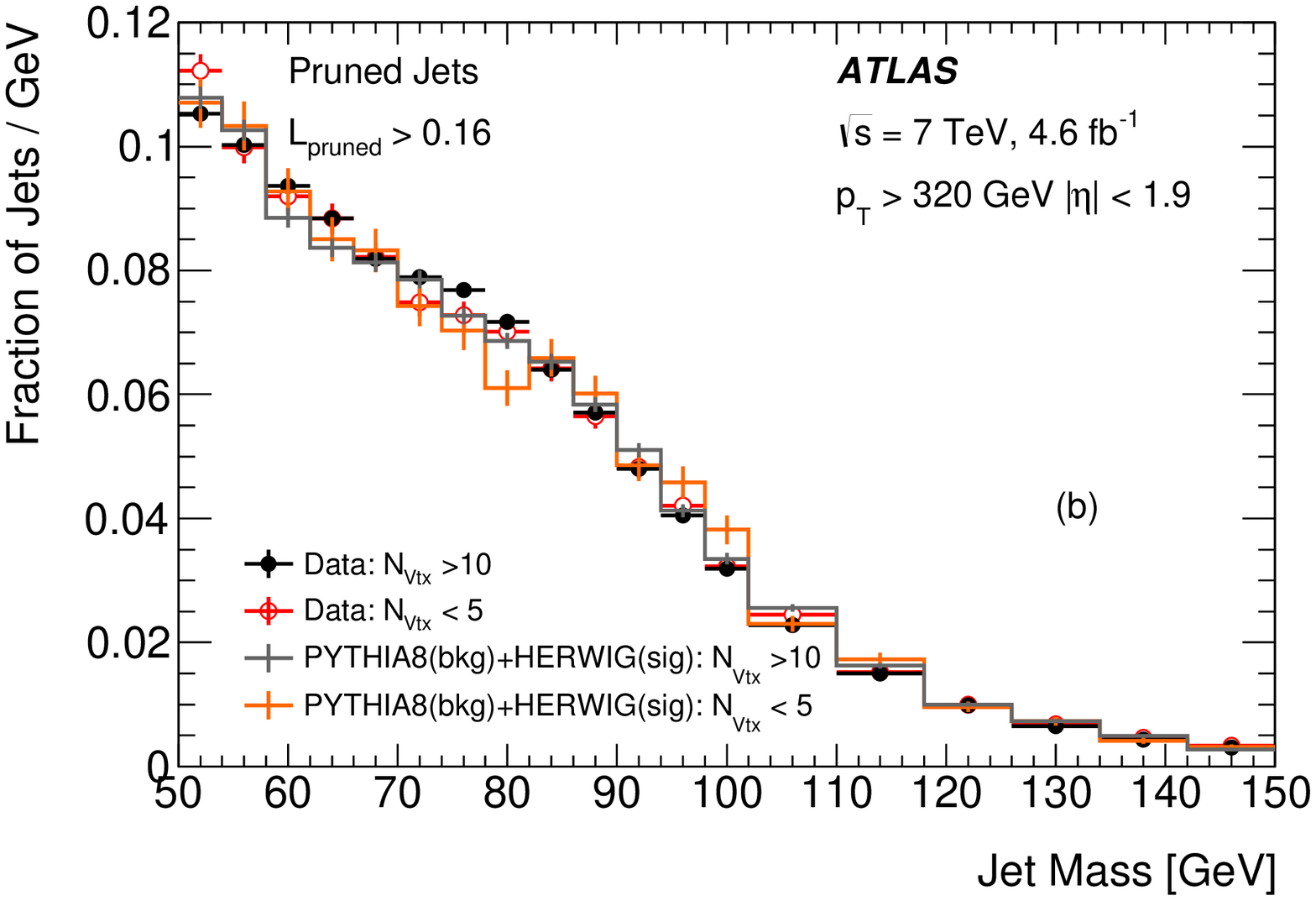}\\
\includegraphics[width=0.45\textwidth] {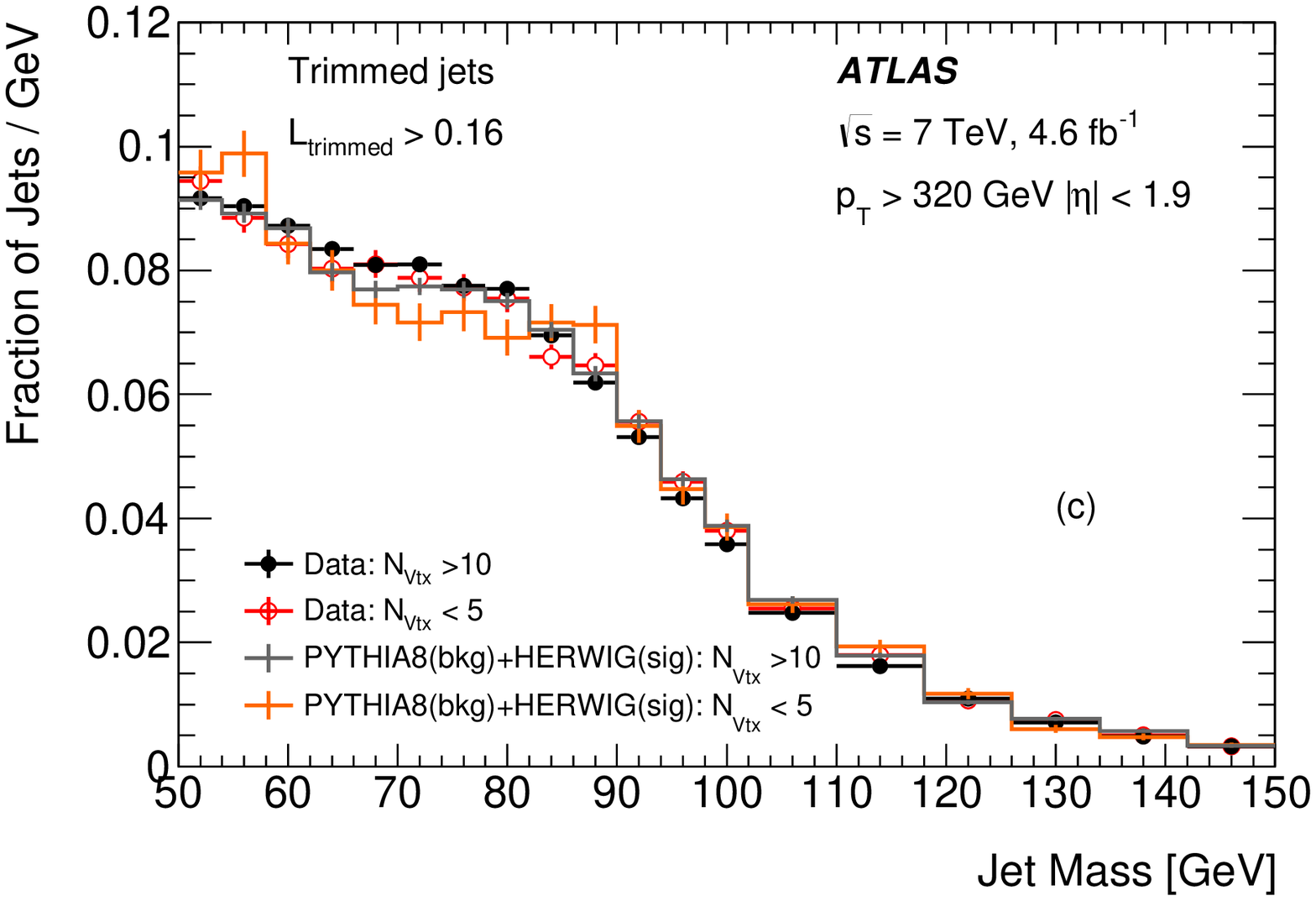} 
\includegraphics[width=0.45\textwidth] {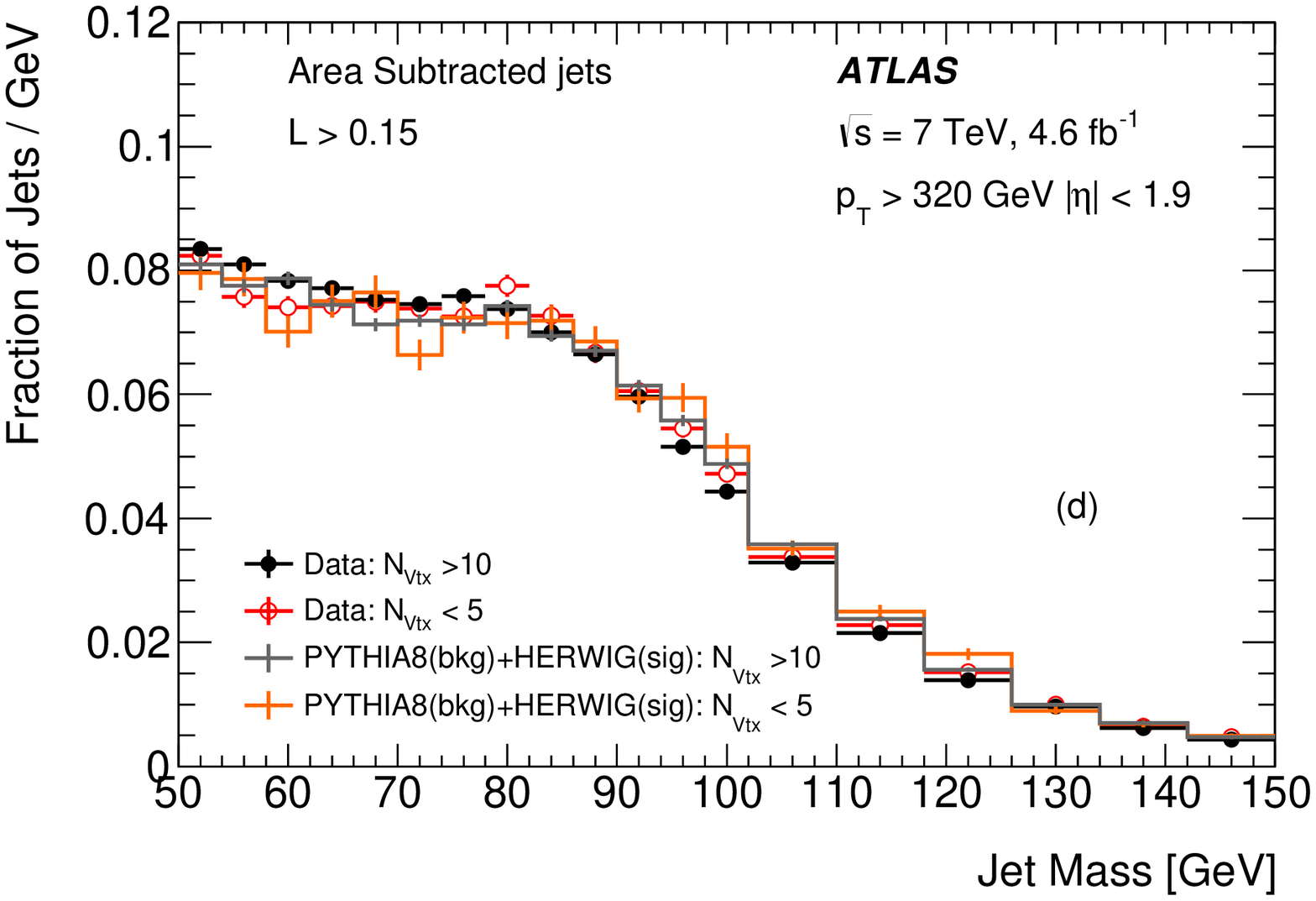}
 \caption{The jet mass distributions for low-pileup ($N_{\textrm{vtx}} < 5$) and high-pileup ($N_{\textrm{vtx}} > 10$) 
conditions in MC (\PYTHIAEIGHT for background plus \HERWIG for signal) and data, where 
$N_{\textrm{vtx}}$ is defined as the number of reconstructed collision vertices in an event. 
The cases without grooming, after applying pruning, trimming and area subtraction are shown. 
Here, $N_{\textrm{vtx}}$ is defined as the number of reconstructed primary vertices in the event.
The uncertainties are statistical only.}
 \label{fig:pileupnogroom}
\end{center}
\end{figure}

Figure~\ref{fig:pileupnogroom}(a) shows the effect of 
pileup on the mass distributions of the ungroomed jets in data and simulation. Events with 
fewer than five reconstructed collision vertices are defined as low-pileup ($N_{\textrm{vtx}} < 5$), events 
with more than ten as high-pileup ($N_{\textrm{vtx}} > 10$). The low- and high-pileup distributions differ 
significantly. The effect of pileup in data is well described by the simulation.

Figure~\ref{fig:pileupnogroom}(b)--~\ref{fig:pileupnogroom}(d) show the same distributions for the pruned, trimmed and area
subtracted jet samples. The mass distributions after grooming are significantly less sensitive to pileup. 
The low- and high-pileup distributions for groomed jets are nearly identical both in data and simulation. 

\section{Conclusion}
This paper presents a measurement of the production cross-section of a hadronically decaying 
boosted $W$ or $Z$ boson with transverse momentum $p_{\rm T}>320\,\gev$ and 
pseudorapidity $|\eta|<1.9$ in \pp  collisions at a centre-of-mass energy of $7\,\tev$ 
with the ATLAS detector at the LHC. The measurement is performed by reconstructing boosted $W$ and $Z$ 
bosons in single jets. The reconstructed jet mass is used to identify the $W$ and $Z$ bosons and 
a jet substructure method based on energy cluster information in the jet centre-of-mass frame 
is used to suppress the large multi-jet background. The measured cross-section is:
\begin{align}
\sigma_{\WplusZ} =
8.5 \pm 0.8\,\stat \pm 1.5\,\syst\,{\rm pb}.\nonumber
\end{align}
The measured value is found to be in agreement with the theoretical prediction for
the same kinematic range of $\sigma_{W+Z}= 5.1\pm 0.5\,{\rm pb}$, obtained from the NLO QCD 
calculation, within $2\,\sigma$. The total uncertainty in the measured 
cross-section is of the same order of magnitude as the uncertainties in measurements 
performed with leptonic decay channels for a similar kinematic region~\cite{Aad:2012en, Aad:2013ysa, Chatrchyan:2011wt}.

The performance of jet grooming techniques has been studied in the context of this analysis.
With comparable cuts on the likelihood ratio and no attempt to optimize the analysis for groomed 
jets, the signal significance is similar for groomed and ungroomed jets. Jet grooming significantly 
reduces the sensitivity to pileup, which will be important in later data taking at the LHC, where 
much higher pileup is expected. In general, the effects of grooming are reasonably well described 
with \PYTHIAEIGHT QCD jet Monte Carlo simulation and \HERWIG \WZ signal Monte Carlo simulation.

\ack


\section{Acknowledgements}

We thank CERN for the very successful operation of the LHC, as well as the
support staff from our institutions without whom ATLAS could not be
operated efficiently.

We acknowledge the support of ANPCyT, Argentina; YerPhI, Armenia; ARC,
Australia; BMWF and FWF, Austria; ANAS, Azerbaijan; SSTC, Belarus; CNPq and FAPESP,
Brazil; NSERC, NRC and CFI, Canada; CERN; CONICYT, Chile; CAS, MOST and NSFC,
China; COLCIENCIAS, Colombia; MSMT CR, MPO CR and VSC CR, Czech Republic;
DNRF, DNSRC and Lundbeck Foundation, Denmark; EPLANET, ERC and NSRF, European Union;
IN2P3-CNRS, CEA-DSM/IRFU, France; GNSF, Georgia; BMBF, DFG, HGF, MPG and AvH
Foundation, Germany; GSRT and NSRF, Greece; ISF, MINERVA, GIF, I-CORE and Benoziyo Center,
Israel; INFN, Italy; MEXT and JSPS, Japan; CNRST, Morocco; FOM and NWO,
Netherlands; BRF and RCN, Norway; MNiSW and NCN, Poland; GRICES and FCT, Portugal; MNE/IFA, Romania; MES of Russia and ROSATOM, Russian Federation; JINR; MSTD,
Serbia; MSSR, Slovakia; ARRS and MIZ\v{S}, Slovenia; DST/NRF, South Africa;
MINECO, Spain; SRC and Wallenberg Foundation, Sweden; SER, SNSF and Cantons of
Bern and Geneva, Switzerland; NSC, Taiwan; TAEK, Turkey; STFC, the Royal
Society and Leverhulme Trust, United Kingdom; DOE and NSF, United States of
America.

The crucial computing support from all WLCG partners is acknowledged
gratefully, in particular from CERN and the ATLAS Tier-1 facilities at
TRIUMF (Canada), NDGF (Denmark, Norway, Sweden), CC-IN2P3 (France),
KIT/GridKA (Germany), INFN-CNAF (Italy), NL-T1 (Netherlands), PIC (Spain),
ASGC (Taiwan), RAL (UK) and BNL (USA) and in the Tier-2 facilities
worldwide.

\clearpage
\appendix
\section{Definitions of jet-shape variables}\label{sec:app}
In the following the definitions of the jet-shape variables used in the \WplusZ cross-section 
measurement are given. 
\begin{itemize}
\item Thrust minor: the thrust axis~\cite{Brandt:1964sa,Farhi:1977sg} of a jet in its centre-of-mass 
frame, $\hat{T}$, is defined as the direction which maximizes the sum of the 
longitudinal momenta of the energy clusters.
The thrust minor~\cite{Brandt:1964sa,Farhi:1977sg}, $T_{\rm min}$, 
is related to this direction and is defined as
\begin{equation}
T_{\rm min}=\frac{\sum_i |\vec{p}_i\times \hat{T}|}{\sum_i |\vec{p}_i|},
\end{equation}
where  $\vec{p}_i$ are the momenta of the energy clusters in the jet rest frame. 
$T_{\rm min}=0$ corresponds to a highly directional distribution of the energy clusters,
and $T_{\rm min}= 0.5$ corresponds to an isotropic distribution.
\item Sphericity: the sphericity tensor~\cite{Bjorken:1969wi} is defined as
\begin{equation}
S^{\alpha\beta}=\frac{\sum_i p^\alpha_i p^\beta_i}{\sum_i |\vec{p}_i|^2},
\end{equation}
where $\alpha$ and $\beta$ correspond to the $x$, $y$ and $z$ components  of the 
momenta of the energy clusters in the jet rest frame. By standard diagonalization 
of $S^{\alpha \beta}$ one may find three eigenvalues $\lambda_1 \geq \lambda_2 \geq \lambda_3$, 
with $\lambda_1 + \lambda_2 + \lambda_3 = 1$. The sphericity is then defined as \
\begin{equation} S = \frac{3}{2} \, (\lambda_2 + \lambda_3).
\end{equation}
Sphericity is a measure of the summed  squares of transverse momenta of all the energy 
clusters with respect to the jet axis. By construction $0 \le S \le 1$. A jet with two back-to-back 
subjets in its rest frame has $S=0$, and $S=1$ indicates an isotropic distribution of the
energy clusters.
\item Aplanarity: the aplanarity~\cite{Bjorken:1969wi} is defined as
\begin{equation}
A=\frac{3\lambda_3}{2},
\end{equation}
and is constrained to the range $0 \leq A \leq \frac{1}{2}$.
A highly directional distribution of the energy clusters has $A=0$, and $A= 0.5$ 
corresponds to an isotropic distribution.
\end{itemize}

\clearpage
\section{Jet grooming algorithms}\label{sec:appC}
The three algorithms studied in this paper behave in different ways with respect to 
the treatment of the constituents of the \WZ jet candidates. The pruning~\cite{Ellis:2009me} and trimming~\cite{Krohn:2009th} algorithms start from 
the constituents of the anti-$k_t$ jets and remove those with kinematic characteristics 
compatible with coming from soft radiation. These modified lists of jet constituents  are used 
to recalculate the jet-shape variables. 
In the trimming algorithm the jet is clustered in subjets, and those with small 
transverse energies are removed. Its main parameters are $R_{\textrm{sub}}$, defined as 
the radius parameter of the subjets obtained after reclustering the jet constituents (using 
the $k_t$ algorithm), and $f_{\textrm{cut}}$, corresponding to the minimum fraction of the 
initial jet transverse momentum carried by the subjets that are retained.
The pruning algorithm uses an iterative jet reclustering method with parameter 
$R_{\textrm{cut}}$, defined as the maximum allowed separation between the subjet 
and the jet, in order to remove large-angle radiation. Denoting by $p_i$ the 
transverse momentum of a subjet, and $p$ that of the jet, only the subjets
with fractional transverse momentum $p_i/p$ larger than a 
parameter $z_{\textrm{cut}}$ are retained. Several possible choices of parameters 
were tested in this study; 
among them the values \{$R_{\textrm{sub}}$ = 0.2, 
$f_{\textrm{cut}}$ = 0.03\} for trimming, and \{$R_{\textrm{cut}}$ = 0.3 and 
$z_{\textrm{cut}}$ = 0.02\} for pruning
were selected, since they give
good discrimination between \WZ jet signal and background with minimal modification 
of the jet mass distribution for the background.

The contributions to jet energies from the underlying event and pileup have large 
fluctuations from event to event. The jet area subtraction technique \cite{areasub} is 
based on the idea that these contributions can be determined on an event-by-event 
basis from all jets in the event. For each event, the distribution of transverse 
energy densities is calculated  from all jets with  $|\eta|<2.1$. 
The transverse energy density of a jet is defined as the ratio of its transverse 
energy to its area. The median of this distribution is taken as an estimate of the energy 
density of the pileup and the underlying event. For each jet, the jet transverse energy 
is corrected by subtracting the product of the transverse energy density and the jet 
area from it. The technique results in a modification of the jet four-momentum, 
including the jet mass. The jet area is determined with the ``active'' area calculation technique~\cite{areasub},
where a large number of infinitely soft particles, so-called ghosts, distributed evenly in the 
($\eta$, $\phi$) plane are included in the jet clustering.\footnote{A second method known as the ``Voronoi'' area technique was found to yield very similar results.} 
The jet area is determined from the number of ghosts that are clustered in the jet.

\newpage
\bibliographystyle{99}  
\bibliography{paper_boostWZ}   

\providecommand{\href}[2]{#2}\begingroup\raggedright\begin{thebibliography}{10}

\bibitem{Chen:2011ah}
C.~Chen, {\em {New approach to identifying boosted hadronically-decaying
  particles using jet substructure in its center-of-mass frame}\/},
  \href{http://dx.doi.org/10.1103/PhysRevD.85.034007}{Phys. Rev. {\bf D85}
  (2012)  034007},
[\href{http://arxiv.org/abs/1112.2567}{{\tt arXiv:1112.2567}}].


\bibitem{Aad:2012en}
{ATLAS} Collaboration, {\em {Study of jets produced in association with a $W$
  boson in $pp$ collisions at $\sqrt{s}=7$ TeV with the ATLAS detector}\/},
  \href{http://dx.doi.org/10.1103/PhysRevD.85.092002}{Phys. Rev. {\bf D85}
  (2012)  092002},
[\href{http://arxiv.org/abs/1201.1276}{{\tt arXiv:1201.1276}}].

\bibitem{Aad:2013ysa}
{ATLAS} Collaboration, {\em {Measurement of the production cross section of
  jets in association with a Z boson in $pp$ collisions at $\sqrt{s} = 7$ TeV
  with the ATLAS detector}\/},
  \href{http://dx.doi.org/10.1007/JHEP07(2013)032}{JHEP {\bf 1307} (2013)
  032},
[\href{http://arxiv.org/abs/1304.7098}{{\tt arXiv:1304.7098}}].

\bibitem{Chatrchyan:2011wt}
{CMS} Collaboration, {\em {Measurement of the rapidity and transverse momentum
  distributions of $Z$ bosons in $pp$ Collisions at $\sqrt{s}=7$ TeV}\/},
  \href{http://dx.doi.org/10.1103/PhysRevD.85.032002}{Phys. Rev. {\bf D85}
  (2012)  032002},
[\href{http://arxiv.org/abs/1110.4973}{{\tt arXiv:1110.4973}}].

\bibitem{Aad:2014bla}
{ATLAS} Collaboration, {\em {Measurement of the cross section of high
  transverse momentum $Z\rightarrow b\bar{b}$ production in proton--proton
  collisions at $\sqrt{s}=8$ TeV with the ATLAS Detector}\/},
[\href{http://arxiv.org/abs/1404.7042}{{\tt arXiv:1404.7042}} (submitted to Phys. Lett. B)].

\bibitem{ATLAS:2012am}
{ATLAS} Collaboration, {\em {Jet mass and substructure of inclusive jets in
  $\sqrt{s}=7$ TeV $pp$ collisions with the ATLAS experiment}\/},
  \href{http://dx.doi.org/10.1007/JHEP05(2012)128}{JHEP {\bf 1205} (2012)
  128},
[\href{http://arxiv.org/abs/1203.4606}{{\tt arXiv:1203.4606}}].

\bibitem{Aad:2013gja}
{ATLAS} Collaboration, {\em {Performance of jet substructure techniques for
  large-$R$ jets in proton-proton collisions at $\sqrt{s}$ = 7 TeV using the
  ATLAS detector}\/},  \href{http://dx.doi.org/10.1007/JHEP09(2013)076}{JHEP
  {\bf 1309} (2013)  076},
[\href{http://arxiv.org/abs/1306.4945}{{\tt arXiv:1306.4945}}].

\bibitem{Chatrchyan:2012ku}
{CMS} Collaboration, {\em {Search for anomalous $t\bar{t}$ production in the
  highly-boosted all-hadronic final state}\/},
  \href{http://dx.doi.org/10.1007/JHEP09(2012)029}{JHEP {\bf 1209} (2012)
  029},
[\href{http://arxiv.org/abs/1204.2488}{{\tt arXiv:1204.2488}}].

\bibitem{Aad:2008zzm}
{ATLAS} Collaboration, {\em {The ATLAS Experiment at the CERN Large Hadron
  Collider}\/},
\href{http://dx.doi.org/10.1088/1748-0221/3/08/S08003}{JINST {\bf 3} (2008)
  S08003}.

\bibitem{Campbell:1999ah}
J.~M. Campbell and R.~K. Ellis, {\em {Update on vector boson pair production at
  hadron colliders}\/},
  \href{http://dx.doi.org/10.1103/PhysRevD.60.113006}{Phys. Rev. {\bf D60}
  (1999)  113006},
[\href{http://arxiv.org/abs/hep-ph/9905386}{{\tt arXiv:hep-ph/9905386}}].

\bibitem{Lai:2010vv}
H.-L. Lai et al., {\em {New parton distributions for collider physics}\/},
  \href{http://dx.doi.org/10.1103/PhysRevD.82.074024}{Phys. Rev. {\bf D82}
  (2010)  074024},
[\href{http://arxiv.org/abs/1007.2241}{{\tt arXiv:1007.2241}}].

\bibitem{Beringer:1900zz}
{Particle~Data~Group}, J.~Beringer, et al., {\em {Review of Particle Physics
  (RPP)}\/},
\href{http://dx.doi.org/10.1103/PhysRevD.86.010001}{Phys. Rev. {\bf D86} (2012)
   010001}.

\bibitem{Aad:2013ucp}
{ATLAS} Collaboration, {\em {Improved luminosity determination in pp collisions
  at $\sqrt{s} = 7$~TeV using the ATLAS detector at the LHC}\/},
  \href{http://dx.doi.org/10.1140/epjc/s10052-013-2518-3}{Eur. Phys. J. {\bf
  C73} (2013)  2518},
[\href{http://arxiv.org/abs/1302.4393}{{\tt arXiv:1302.4393}}].

\bibitem{Corcella:2000bw}
G.~Corcella et al., {\em {HERWIG 6: An Event generator for hadron emission
  reactions with interfering gluons (including supersymmetric processes)}\/},
  \href{http://dx.doi.org/10.1088/1126-6708/2001/01/010}{JHEP {\bf 0101} (2001)
   010},
[\href{http://arxiv.org/abs/hep-ph/0011363}{{\tt arXiv:hep-ph/0011363}}].

\bibitem{Butterworth:1996zw}
J.~M. Butterworth, J.~R. Forshaw, and M.~H. Seymour, {\em {Multiparton
  interactions in photoproduction at HERA}\/},
  \href{http://dx.doi.org/10.1007/s002880050286}{Z. Phys. {\bf C72} (1996)
  637},
[\href{http://arxiv.org/abs/hep-ph/9601371}{{\tt arXiv:hep-ph/9601371}}].

\bibitem{Sherstnev:2007nd}
A.~Sherstnev and R.~S. Thorne, {\em {Parton Distributions for LO
  Generators}\/},  \href{http://dx.doi.org/10.1140/epjc/s10052-008-0610-x}{Eur.
  Phys. J. {\bf C55} (2008)  553},
[\href{http://arxiv.org/abs/0711.2473}{{\tt arXiv:0711.2473 [hep-ph]}}].

\bibitem{Sherstnev:2008dm}
A.~Sherstnev and R.~S. Thorne, {\em {Different PDF approximations useful for LO
  Monte Carlo generators}\/},
[\href{http://arxiv.org/abs/0807.2132}{{\tt arXiv:0807.2132}}].

\bibitem{Sjostrand:2007gs}
T.~Sjostrand, S.~Mrenna, and P.~Z. Skands, {\em {A Brief Introduction to PYTHIA
  8.1}\/},  \href{http://dx.doi.org/10.1016/j.cpc.2008.01.036}{Comput. Phys.
  Commun. {\bf 178} (2008)  852},
[\href{http://arxiv.org/abs/0710.3820}{{\tt arXiv:0710.3820}}].

\bibitem{Sjostrand:2006za}
T.~Sjostrand, S.~Mrenna, and P.~Z. Skands, {\em {PYTHIA 6.4 physics and
  manual}\/},  \href{http://dx.doi.org/10.1088/1126-6708/2006/05/026}{JHEP {\bf
  0605} (2006)  026},
[\href{http://arxiv.org/abs/hep-ph/0603175}{{\tt arXiv:hep-ph/0603175}}].

\bibitem{Aad:2010ac}
{ATLAS} Collaboration, {\em {Charged-particle multiplicities in pp interactions
  measured with the ATLAS detector at the LHC}\/},
  \href{http://dx.doi.org/10.1088/1367-2630/13/5/053033}{New J. Phys. {\bf 13}
  (2011)  053033},
[\href{http://arxiv.org/abs/1012.5104}{{\tt arXiv:1012.5104}}].

\bibitem{Pumplin:2002vw}
J.~Pumplin et al., {\em {New generation of parton distributions with
  uncertainties from global QCD analysis}\/},
  \href{http://dx.doi.org/10.1088/1126-6708/2002/07/012}{JHEP {\bf 0207} (2002)
   012},
[\href{http://arxiv.org/abs/hep-ph/0201195}{{\tt arXiv:hep-ph/0201195}}].

\bibitem{Bahr:2008pv}
M.~Bahr et al., {\em {Herwig++ physics and manual}\/},
  \href{http://dx.doi.org/10.1140/epjc/s10052-008-0798-9}{Eur. Phys. J. {\bf
  C58} (2008)  639},
[\href{http://arxiv.org/abs/0803.0883}{{\tt arXiv:0803.0883}}].

\bibitem{Skands:2010ak}
P.~Z. Skands, {\em {Tuning Monte Carlo generators: The Perugia tunes}\/},
  \href{http://dx.doi.org/10.1103/PhysRevD.82.074018}{Phys. Rev. {\bf D82}
  (2010)  074018},
[\href{http://arxiv.org/abs/1005.3457}{{\tt arXiv:1005.3457}}].

\bibitem{Alioli:2010xd}
S.~Alioli, P.~Nason, C.~Oleari, and E.~Re, {\em {A general framework for
  implementing NLO calculations in shower Monte Carlo programs: the POWHEG
  BOX}\/},  \href{http://dx.doi.org/10.1007/JHEP06(2010)043}{JHEP {\bf 1006}
  (2010)  043},
[\href{http://arxiv.org/abs/1002.2581}{{\tt arXiv:1002.2581}}].

\bibitem{Alioli:2010xa}
S.~Alioli, K.~Hamilton, P.~Nason, C.~Oleari, and E.~Re, {\em {Jet pair
  production in POWHEG}\/},
  \href{http://dx.doi.org/10.1007/JHEP04(2011)081}{JHEP {\bf 1104} (2011)
  081},
[\href{http://arxiv.org/abs/1012.3380}{{\tt arXiv:1012.3380}}].

\bibitem{Frixione:2002ik}
S.~Frixione and B.~R. Webber, {\em {Matching NLO QCD computations and parton
  shower simulations}\/},
  \href{http://dx.doi.org/10.1088/1126-6708/2002/06/029}{JHEP {\bf 0206} (2002)
   029},
[\href{http://arxiv.org/abs/hep-ph/0204244}{{\tt arXiv:hep-ph/0204244}}].

\bibitem{Czakon:2013goa}
M.~Czakon, P.~Fiedler, and A.~Mitov, {\em {Total Top-Quark Pair-Production
  Cross Section at Hadron Colliders Through $O(\alpha^{4}_{S})$}\/},
  \href{http://dx.doi.org/10.1103/PhysRevLett.110.252004}{Phys. Rev. Lett. {\bf
  110} (2013)  252004},
[\href{http://arxiv.org/abs/1303.6254}{{\tt arXiv:1303.6254}}].

\bibitem{Agostinelli:2002hh}
{GEANT4} Collaboration, S.~Agostinelli et al., {\em {GEANT4---a Simulation
  toolkit}\/},
\href{http://dx.doi.org/10.1016/S0168-9002(03)01368-8}{Nucl. Instrum. Meth.
  {\bf A506} (2003)  250}.

\bibitem{Aad:2010ah}
{ATLAS} Collaboration, {\em {The ATLAS Simulation Infrastructure}\/},
  \href{http://dx.doi.org/10.1140/epjc/s10052-010-1429-9}{Eur. Phys. J. {\bf
  C70} (2010)  823},
[\href{http://arxiv.org/abs/1005.4568}{{\tt arXiv:1005.4568}}].

\bibitem{Cacciari:2008gp}
M.~Cacciari, G.~P. Salam, and G.~Soyez, {\em {The anti-$k_t$ jet clustering
  algorithm}\/},  \href{http://dx.doi.org/10.1088/1126-6708/2008/04/063}{JHEP
  {\bf 0804} (2008)  063},
[\href{http://arxiv.org/abs/0802.1189}{{\tt arXiv:0802.1189}}].

\bibitem{Aad:2010ad}
{ATLAS} Collaboration, {\em {Measurement of inclusive jet and dijet cross
  sections in proton-proton collisions at 7 TeV centre-of-mass energy with the
  ATLAS detector}\/},
  \href{http://dx.doi.org/10.1140/epjc/s10052-010-1512-2}{Eur. Phys. J. {\bf
  C71} (2011)  1512},
[\href{http://arxiv.org/abs/1009.5908}{{\tt arXiv:1009.5908}}].

\bibitem{Aad:2012vm}
{ATLAS} Collaboration, {\em {Single hadron response measurement and calorimeter
  jet energy scale uncertainty with the ATLAS detector at the LHC}\/},
  \href{http://dx.doi.org/10.1140/epjc/s10052-013-2305-1}{Eur. Phys. J. {\bf
  C73} (2013)  2305},
[\href{http://arxiv.org/abs/1203.1302}{{\tt arXiv:1203.1302}}].

\bibitem{Aad:2014bia}
{ATLAS} Collaboration, {\em {Jet energy measurement and its systematic
  uncertainty in proton--proton collisions at $\sqrt{s}=7$~TeV with the ATLAS
  detector}\/},
[\href{http://arxiv.org/abs/1406.0076}{{\tt arXiv:1406.0076}}].

\bibitem{Brandt:1964sa}
S.~Brandt, Ch.~Peyrou, R.~Sosnowski, and A.~Wroblewski, {\em {The principal axis
  of jets --- an attempt to analyze high-energy collisions as two-body
  processes}\/},
Phys. Lett. {\bf 12} (1964)  57.

\bibitem{Farhi:1977sg}
E.~Farhi, {\em {Quantum Chromodynamics Test for Jets}\/},
\href{http://dx.doi.org/10.1103/PhysRevLett.39.1587}{Phys. Rev. Lett. {\bf 39}
  (1977)  1587}.

\bibitem{Bjorken:1969wi}
J.~D. Bjorken and S.~J. Brodsky, {\em {Statistical Model for Electron-Positron
  Annihilation Into Hadrons}\/},
\href{http://dx.doi.org/10.1103/PhysRevD.1.1416}{Phys. Rev. {\bf D1} (1970)
  1416}.

\bibitem{Aad:2011he}
{ATLAS} Collaboration, {\em {Jet energy measurement with the ATLAS detector in
  proton-proton collisions at $\sqrt{s}=7$ TeV}\/},
  \href{http://dx.doi.org/10.1140/epjc/s10052-013-2304-2}{Eur. Phys. J. {\bf
  C73} (2013)  2304},
[\href{http://arxiv.org/abs/1112.6426}{{\tt arXiv:1112.6426}}].

\bibitem{Aad:2012ag}
{ATLAS} Collaboration, {\em {Jet energy resolution in proton-proton collisions
  at $\sqrt{s}=7$ TeV recorded in 2010 with the ATLAS detector}\/},
  \href{http://dx.doi.org/10.1140/epjc/s10052-013-2306-0}{Eur. Phys. J. {\bf
  C73} (2013)  2306},
[\href{http://arxiv.org/abs/1210.6210}{{\tt arXiv:1210.6210}}].

\bibitem{Andersson:1992iq}
B.~Andersson, G.~Gustafson, and H.~Pi, {\em {The FRITIOF model for very
  high-energy hadronic collisions}\/},
\href{http://dx.doi.org/10.1007/BF01474343}{Z. Phys. {\bf C57} (1993)  485}.

\bibitem{Heikkinen:2003sc}
A.~Heikkinen, N.~Stepanov, and J.~P. Wellisch, {\em {Bertini intranuclear
  cascade implementation in GEANT4}\/},  eConf {\bf C0303241} (2003)  MOMT008,
[\href{http://arxiv.org/abs/nucl-th/0306008}{{\tt arXiv:nucl-th/0306008}}].

\bibitem{Amelin:1989wa}
N.~S. Amelin, K.~K. Gudima, and V.~D. Toneev, {\em {Quark - Gluon String Model
  and Ultrarelativistic Heavy Ion Interactions. (In German)}\/},
Sov. J. Nucl. Phys. {\bf 51} (1990)  327.

\bibitem{Ellis:2009me}
S.~D. Ellis, C.~K. Vermilion, and J.~R. Walsh, {\em {Recombination algorithms
  and jet substructure: Pruning as a tool for heavy particle searches}\/},
  \href{http://dx.doi.org/10.1103/PhysRevD.81.094023}{Phys. Rev. {\bf D81}
  (2010)  094023},
[\href{http://arxiv.org/abs/0912.0033}{{\tt arXiv:0912.0033}}].

\bibitem{Krohn:2009th}
D.~Krohn, J.~Thaler, and L.-T. Wang, {\em {Jet trimming}\/},
  \href{http://dx.doi.org/10.1007/JHEP02(2010)084}{JHEP {\bf 1002} (2010)
  084},
[\href{http://arxiv.org/abs/0912.1342}{{\tt arXiv:0912.1342}}].

\bibitem{areasub}
M.~Cacciari and G.~P. Salam, {\em {Pileup subtraction using jet areas}\/},
  \href{http://dx.doi.org/10.1016/j.physletb.2007.09.077}{Phys. Lett. B {\bf
  659} (2008)  119},
[\href{http://arxiv.org/abs/0707.1378}{{\tt arXiv:0707.1378}}].

\end{thebibliography}\endgroup

\clearpage

\begin{flushleft}
{\Large The ATLAS Collaboration}

\bigskip

G.~Aad$^{\rm 84}$,
T.~Abajyan$^{\rm 21}$,
B.~Abbott$^{\rm 112}$,
J.~Abdallah$^{\rm 152}$,
S.~Abdel~Khalek$^{\rm 116}$,
O.~Abdinov$^{\rm 11}$,
R.~Aben$^{\rm 106}$,
B.~Abi$^{\rm 113}$,
M.~Abolins$^{\rm 89}$,
O.S.~AbouZeid$^{\rm 159}$
% ...
\bigskip
\\
% ...
$^{11}$ Institute of Physics, Azerbaijan Academy of Sciences, Baku, Azerbaijan\\
% ...
$^{21}$ Physikalisches Institut, University of Bonn, Bonn, Germany\\
% ...
$^{84}$ CPPM, Aix-Marseille Universit{\'e} and CNRS/IN2P3, Marseille, France\\
% ...
$^{89}$ Department of Physics and Astronomy, Michigan State University, East Lansing MI, United States of America\\
% ...
$^{106}$ Nikhef National Institute for Subatomic Physics and University of Amsterdam, Amsterdam, Netherlands\\
% ...
$^{112}$ Homer L. Dodge Department of Physics and Astronomy, University of Oklahoma, Norman OK, United States of America\\
$^{113}$ Department of Physics, Oklahoma State University, Stillwater OK, United States of America\\
% ...
$^{116}$ LAL, Universit{\'e} Paris-Sud and CNRS/IN2P3, Orsay, France\\
% ...
$^{152}$ Institute of Physics, Academia Sinica, Taipei, Taiwan\\
% ...
$^{159}$ Department of Physics, University of Toronto, Toronto ON, Canada\\
% ...
\end{flushleft}

\end{document}